\documentclass[symmetry,review,submit,moreauthors,pdftex,12pt,a4paper]{mdpi} 
\lastpage{x}
\doinum{10.3390/------}
\pubvolume{xx}
\pubyear{2010}
\history{Received: xx / Accepted: xx / Published: xx}
\pdfoutput=1

\newcommand{\be}{\begin{equation}}
\newcommand{\ee}{\end{equation}}
\newcommand{\bea}{\begin{eqnarray}}
\newcommand{\eea}{\end{eqnarray}}
\newcommand{\ba}{\begin{array}}
\newcommand{\ea}{\end{array}}

\usepackage{amssymb,amsmath}
\usepackage{graphicx}
\usepackage{epsfig}
\usepackage{caption}
\Title{Complex Networks and Symmetry I: A Review}

\Author{Diego Garlaschelli $^{1}$, Franco Ruzzenenti $^{2,\star}$ and Riccardo Basosi $^{3}$}

\address{%
$^{1}$ LEM, Sant'Anna School of Advanced Studies, P.zza Martiri della Libert\`a 33, 56127 Pisa, Italy;\linebreak E-Mail: garlaschelli.diego@gmail.com\\
$^{2}$ Center for the Study of Complex Systems, University of Siena, Via Roma 56, 53100 Siena, Italy\\
$^{3}$ Department of Chemistry, University of Siena, Via Aldo Moro
1, 53100 Siena, Italy; \linebreak E-Mail:basosi@unisi.it}

\corres{E-Mail: ruzzenenti@unisi.it; \linebreak Tel.: +39-0577-234240; Fax: +39-0577-234239.}

\abstract{In this review we establish various connections between
complex networks and symmetry. While special types of symmetries
(e.g., automorphisms) are studied in detail within discrete
mathematics for particular classes of deterministic graphs, the
analysis of more general symmetries in real complex networks is far
less developed. We argue that real networks, as any entity
characterized by imperfections or errors, necessarily require a
stochastic notion of invariance. We therefore propose a definition
of stochastic symmetry based on graph ensembles and use it to review
the main results of network theory from an unusual perspective. The
results discussed here and in a companion paper show that stochastic
symmetry highlights the most informative topological properties of
real networks, even in noisy situations unaccessible to exact
techniques.}


\begin{document}
\section{Introduction\label{sec_levels}}
In this review and in a companion paper \cite{symmetry2}, we study
several connections between symmetry and network theory. Most
complex systems encountered in a diverse range of domains, from
biology through sociology to technology, consist of networks of
elements (\emph{vertices}) connected together (by \emph{links}, or
\emph{edges}) in an intricate way
\cite{guidosbook,largescalestructure,
dynamicalprocessesoncomplexnetworks,
ecologicalnetworks,internet,networksincellbiology,adaptivenetworks}.
While graph theory started dealing with the mathematical description
of network properties long ago \cite{harary}, only recently massive
datasets about large real-world complex networks have become
available. This allowed an unprecedented activity of data analysis,
which resulted in the establishment of some key `stylized facts'
about the structure of real networks, and motivated an intense
theoretical activity aimed at explaining them.

Surprisingly (at least at the time when this was first observed),
the empirically observed structure of real networks is strikingly
different from what is obtained assuming simple homogeneous
mechanisms of network formation, such as the traditional
Erd\H{o}s-R\'enyi random graph model
\cite{guidosbook,largescalestructure}. In the latter, which will be
an important reference throughout this review, every pair of
vertices has the same probability $p$ to be connected. This
generates homogeneous topological features, such as a constant link
density across the network, and a narrow (binomial) distribution of
the degree $k$ (number of edges reaching a vertex). By contrast,
virtually any real network is found to display a modular structure,
with vertices organized in communities tightly connected internally
and loosely connected to each other, and a broad degree
distribution, typically featuring a power-law tail of the form
$P(k)\propto k^{-\gamma}$. Networks characterized by the latter
property are called \emph{scale-free}.

Besides the purely topological level, networks are also
characterized by heterogeneous link weights. That is, the intensity
of the connections is again broadly distributed, and non-trivially
correlated to the topology. Capturing the richness of the
information encoded in the weighted structure of networks is a hard
task, and the definition of proper weighted structural properties an
open problem \cite{vespy_weighted,myensemble,kertesz_clustering}. \linebreak At
both the topological and the weighted level, many real networks are
also characterized by an intrinsic directionality of their
connections, which again implies that proper quantities must be
introduced and measured in order to fully understand directed
structural patterns
\cite{myreciprocity,mymultispecies,giorgioclustering}.

As an additional level of complexity, dynamical processes generally
take place on networks \cite{dynamicalprocessesoncomplexnetworks}.
Remarkably, the heterogeneous structure of real networks has been
found to determine major deviations from the behavior expected in
the homogeneous case, which is the traditional assumption used to
obtain predictions about the dynamics. As a consequence, most of
these predictions have been shown to be incorrect when applied to
real-world networks. A prototypic example of this discrepancy is
found in models of epidemic disease spreading. When these models are
defined on regular graphs, one finds that the transmission rate must
overcome a finite \emph{epidemic threshold} in order to guarantee
the persistence of an infection. By contrast, on scale-free networks
the value of the epidemic threshold vanishes, implying that a large
class of diseases can escape extinction no matter their transmission
rates, even if extremely low
\cite{dynamicalprocessesoncomplexnetworks}.

Finally, in some networks a feedback is present between the topology
and the dynamics taking place on it. This is the case of adaptive
networks, whose structure changes in response to their dynamical
behavior, which is in turn affected by the structure itself
\cite{adaptivenetworks}. Generally, adaptive networks cannot be
properly understood by studying their topology and their dynamics
separately, as simple models \linebreak show \cite{myselforganized}.

It may appear that, due to the various levels of complexity
encountered in the description of real networks, performing a
symmetry analysis of these highly heterogeneous systems is likely to
lead to a dead end. This is probably the reason why, although
network theory developed very rapidly in recent years
\cite{guidosbook,largescalestructure,dynamicalprocessesoncomplexnetworks}
and established tight connections with many other disciplines
\cite{ecologicalnetworks,internet,networksincellbiology,adaptivenetworks},
its many relations to symmetry concepts have not been made explicit
yet, apart from isolated examples
\cite{symmetry,quotient,redundancy,symmetry_wtw}. On the other hand,
one expects the formation of real networks to be guided by some
organising principle, maybe non-obvious but surely not completely
random, and possibly the result of evolutionary or optimisation
mechanisms. This implies that network structure should encode some
degree of order and symmetry, even if more general and challenging
than the type found in geometrical objects. It is therefore
important to introduce proper definitions of symmetries capturing
the possible forms of organisation of real networks, and enabling a
simplified understanding of the latter.

In this review we explore the connections between real networks and
symmetry in more detail. We show that many of the approaches that
have been proposed to characterize both real and model-generated
networks can be rephrased more firmly in terms of symmetry concepts.
To this end, in Section \ref{sec_transformations} we first clarify
the peculiar notions of symmetry pertinent to real networks, which
(unlike formal graphs studied in discrete mathematics) are always
characterized by errors or imperfections. Then, in Section
\ref{sec_symmetry} we shall establish several connections between
network theory and symmetry. Symmetry will be investigated over a
wide range of invariances related to topological variables. The
empirical result that in real networks some topological properties
tend to distribute in structurally different ways from random
networks, thus emphasizing a complex structure, will be rephrased in
terms of symmetry concepts. Interestingly, \linebreak Section
\ref{sec_symmetry} can be regarded as a brief review of network
theory from the unusual perspective of the symmetry properties of
real networks. Finally, in Section \ref{sec_conclusions} we
summarize our survey of network symmetries. In the companion paper
\cite{symmetry2}, we exploit the concepts developed here to study
stochastic symmetry in great detail in a particular case, and to
address the problem of symmetry breaking in networks.

\section{Types of Graph Symmetries\label{sec_transformations}}
Before proceeding with a review of the empirical symmetries of real
networks, we first distinguish between different notions of
invariance we will be interested in. The mathematical definition of
symmetry of an object is the set of transformations that leave the
properties of the object unchanged. For instance, a straight line of
infinite length is unchanged after displacing it along its own
direction, and a circle is unchanged after rotating it around its
center. Conversely, the transformations involved in the symmetries
of an object can be exploited to define and construct the object
itself: a straight line can be drawn by displacing a point along a
chosen direction, and a circle can be drawn by rotating a point
around a chosen center. In this case, one needs a \emph{unit} (in
both examples, a point) to iterate through the transformation. More
complicated units lead to more complicated objects (for instance,
rotating a whole disk rather than a single point leads to a torus).

\subsection{Discreteness, Permutations, and More General Symmetries}
\begin{figure}[h]
\begin{center}
\includegraphics[width=0.8\textwidth]{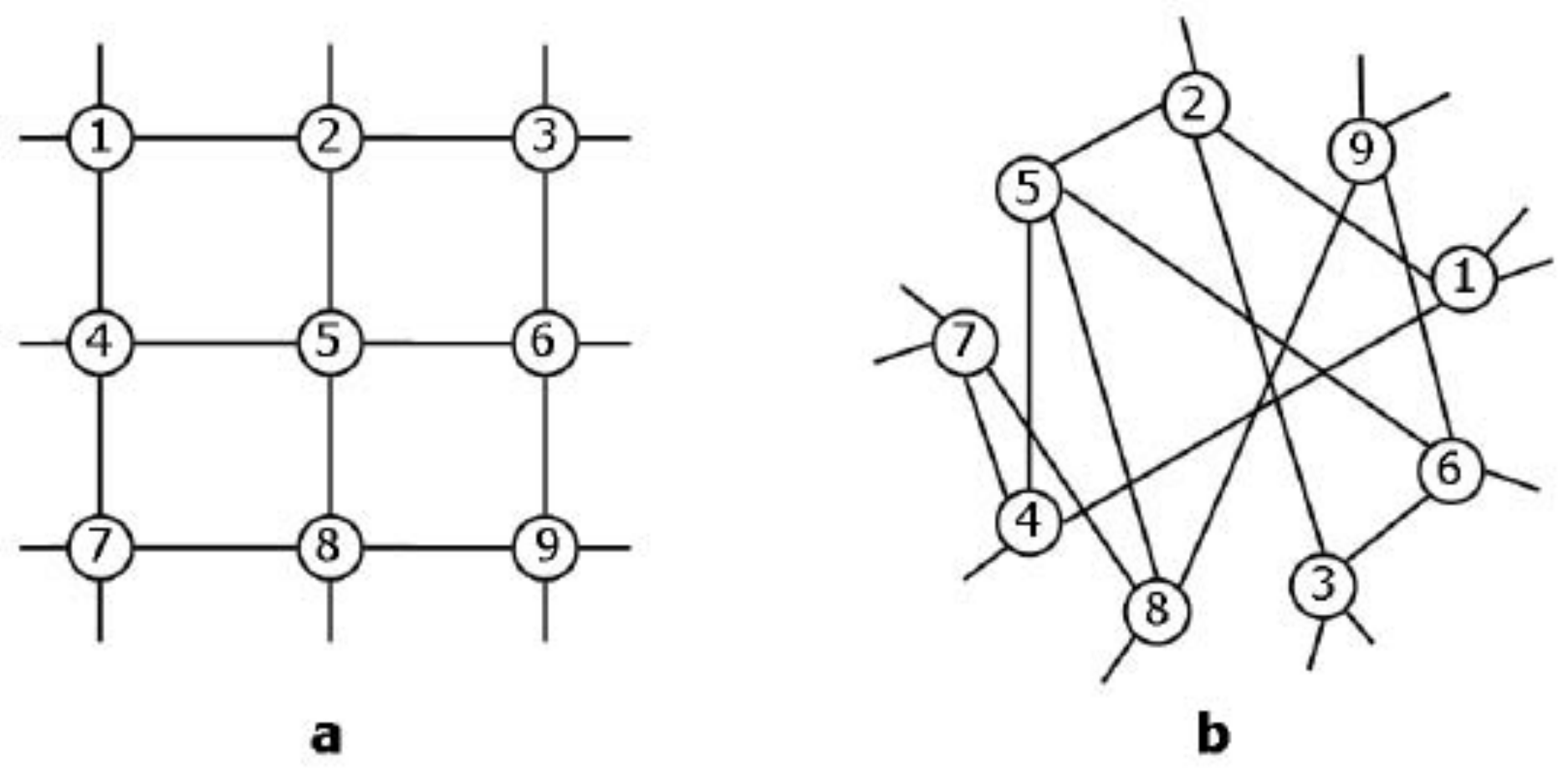}
\end{center}
\caption{A two-dimensional lattice (in principle of infinite size)
constructed by assigning equally spaced planar coordinates to
vertices, and connecting each vertex to its nearest neighbours. (a)
The lattice is visualized by drawing vertices according to their
coordinates in the embedding space. (b) The same graph is drawn by
arbitrarily positioning vertices, irrespective of their coordinates
in the embedding space. Mathematically, the two graphs are
indistinguishable and are therefore characterized by the same
automorphisms (permutations of vertices leading to the same
topology). The knowledge of the vertices' positions [evident in (a)]
directly indicates which are the permutations corresponding to the
symmetries of the graph: only those that map all vertices to their
nearest neighbours in a specified direction. Even if it is natural
to regard such transformations as translations or displacements
(with respect to the embedding space), topologically they are mere
permutations of vertices. \label{fig_embed}} \vspace{-0.5cm}
\end{figure}

In the case of graph symmetries, various considerations are in
order. Firstly, a graph is a discrete object, and therefore the
relevant transformations are discrete and not continuous as in the
previous examples. An example of discrete transformation is the
rotation of a square by an angle of $\pi/2$ radians (or a multiple)
around its center: the square is symmetric under this discrete
rotation but not under one with different angle, or under a
continuous one. Similarly, a lattice (see Figure \ref{fig_embed}) is
symmetric under a discrete displacement by a multiple of the lattice
spacing (which maps all vertices to their nearest neighbours in a
specified direction), but not under one with different length, or
under a continuous one. We will discuss the discrete translational
symmetry of networks in Section \ref{sec_translational}.

Secondly, graphs are topological objects, not geometrical entities:
their properties are independent on the positions of vertices in
some metric space, even if the graph itself may be the result of
some position-dependent construction rule. Changing the positions
(and sizes) of vertices, as well as changing the lengths (and
widths) of links, only leads to a different visualisation of the
same graph (see \linebreak Figure \ref{fig_embed}), and has no
effect on the topology of the latter (provided each link remains
attached to the original vertices). Therefore the properties in
terms of which one can check the symmetries of a graph are purely
topological, and the set of transformations involved in such
symmetries are purely relational. Whereas a geometric transformation
(such as a translation or rotation) maps each point in a circle to a
different point determined by its coordinates in the plane, a
topological transformation maps each vertex in a graph to a vertex
determined not by its coordinates but simply by its identity ({\em
i.e.} its label): this transformation is merely a permutation of
vertices. Permutations of vertices leading to the same topology are
the \emph{automorphisms} of a graph, and we will discuss them in
Section \ref{sec_structural}. Nonetheless, if vertices are assigned
coordinates in some embedding space, and the network construction
depends on those coordinates (as for lattices), then the
automorphisms are permutations induced by proper coordinate
transformations (e.g. a translation). This is illustrated  in Figure
\ref{fig_embed}. Similar considerations apply if the graph
construction depends on other properties, rather than positions,
assigned to vertices (we will consider this case explicitly in
Section \ref{sec_properties}). However, in many real-world cases one
only knows the topology of the graph, and not the properties of
vertices. In general, one does not even know whether vertices are
actually assigned properties on which the structure of the network
depends. In this case, all the automorphisms of the graph must be
looked for by enumeration. The possibility that the  complexity of
real networks might be traced back to some simpler description
involving \emph{hidden} variables attached to vertices, whose
transformations may induce symmetries that are not evident \emph{a
priori}, is an important aspect of network research, that we will
discuss in Section \ref{sec_properties}. Therefore the general
problem of graph automorphisms (Section \ref{sec_structural}) can
take different forms depending on the nature of the (possible)
properties inducing the symmetries of a particular network, as the
two examples of translational symmetry \linebreak (Section
\ref{sec_translational}) and permutation of vertex properties
(Section \ref{sec_properties}) show.

Thirdly, graphs may (or may not) exhibit symmetries under
transformations that are not necessarily vertex permutations. An
example is \emph{scale invariance} (Section
\ref{sec_scaleinvariance}), which also applies to self-similar
geometric objects, or \emph{fractals}. In this case, the
transformation is a change of scale in the description of the
system. We will also encounter transformations that drastically
change the topology of a graph, and only preserve some specified
property such as the total number of links or the degrees of all
vertices (Section \ref{sec_equiprobability}). Finally, the
transformations we will consider in Sections \ref{sec_communities}
and \ref{sec_weights} are vertex partitions and edge (rather than
vertex) permutations respectively.

\subsection{Stochastic Symmetry}
As a final important remark we note that, when considering real
networks rather than abstract graphs, one must take into account
that the observed symmetry is in general only approximate. To
illustrate this concept, let us consider the example of a real
object of circular shape. While a perfect circle, as a mathematical
entity, displays an exact rotational symmetry around its center, a
real circle is unavoidably an approximate object, characterized by
small imperfections. If we look for exact rotational symmetry in
real circles, we have to conclude that no real object is circular,
as perfect circles do not exist in reality. The paradox can only be
solved by introducing an approximate notion of rotational symmetry,
{\em i.e.} one where we allow rotated points to fall \emph{nearby}
existing points of the circle. Ultimately, this changes the picture
substantially, since while a perfect circle can only be drawn by a
perfectly rotating point ({\em i.e.} there is a unique trajectory
defining the circle), an imperfect circle can be drawn following
(infinitely) many trajectories. While there is a single perfect
circle of given radius, there are infinite imperfect circles of
given radius (and the definitions of circle and radius themselves
also acquire an approximate meaning).

Thus, while we started investigating the symmetry of a single
object, we naturally end up with a \emph{family} of objects
(containing the original one), all different from each other but
nonetheless characterized by the same approximate symmetry.
Remarkably, to define the symmetry of the single object, we need the
entire family of its variants: while a rotation maps a perfect
circle to itself, it maps an imperfect circle to a different
imperfect circle. In particular, if we assume that a probability is
associated with each approximate object (for instance, if we draw a
circle by adding a small noise term in the radial direction), we end
up with what is known as a \emph{statistical ensemble} of objects.
Objects `closer' to the perfectly symmetric one are assigned larger
probability, and objects deviating from the perfectly symmetric one
by the same amount are equiprobable. A given real object is
symmetric under the transformation considered if it is a
\emph{typical} ({\em i.e.} not unlikely) member of the ensemble
defined by the transformation itself. This also means that, in order
to detect deviations from symmetry in real objects, one needs an
ensemble of imperfectly symmetric objects as a reference or
\emph{null model}. For instance, suppose one is investigating the
properties of a real circle and a real square under rotational
symmetry. If a perfect circle is assumed as the reference for a
rotationally symmetric object, than both the real square and the
real circle will be classified as non-symmetric. By contrast, if an
ensemble of imperfect circles is considered as the null model, then
the real square will still be classified as non-symmetric (since it
is a very unlikely outcome of a circular null model) but the real
circle will now be correctly classified as symmetric.

When applied to networks, the above considerations naturally lead to
the notion of \emph{statistical ensembles of graphs}, {\em i.e.}
families of networks where each graph $G$ is assigned a probability
$P(G)$. We will encounter graph ensembles when considering either
approximate equivalences or null models of real networks. As in the
example above, a graph will be classified as \emph{exactly
symmetric} under a given transformation if it is mapped onto itself
by the transformation  (graph automorphism are an example of exact
vertex permutation symmetry). By contrast, a graph will be
classified as \emph{stochastically symmetric} under a given
transformation if it is a typical member of ({\em i.e.} well
reproduced by) a graph ensemble which is stochastically symmetric
under the same transformation. In the last definition, we consider a
graph ensemble as stochastically symmetric under a transformation if
the latter maps a graph $G_1$ in the ensemble into an equiprobable
graph $G_2$ with $P(G_2)=P(G_1)$. Graph ensembles as null models of
real networks will be introduced and discussed in Section
\ref{sec_equiprobability}, where we will also illustrate in more
detail the idea of stochastic symmetry. We will also show that
stochastic symmetry and entropy are intimately related in graph
ensembles.

\section{Symmetries in Real Networks\label{sec_symmetry}}
Thus there are various possible notions of symmetry one can look for
in networks. In what follows, rather than discussing them in the
order presented above, we follow a more pedagogical ordering, which
allows us to trace the main results of network theory from the
unusual perspective of symmetry. As we will try to elucidate, some
symmetries are generally present in real networks, others are
generally absent, and others are strongly network-dependent and
variably observed.  In some cases, even when a symmetry is present,
it only holds within a limited range. All these situations are
equally important, as they suggest what is relevant and what is not
to plausible formation mechanisms involving a particular network.
Our discussion provides a somewhat unconventional overview of this
problem, and list a few examples (among possibly many more) of
symmetries relevant to networks. Readers interested in a more
comprehensive account of the results of network theory are referred
to the relevant literature
\cite{guidosbook,largescalestructure,dynamicalprocessesoncomplexnetworks,adaptivenetworks}
and to the publications cited in the following text.


\subsection{Translational Symmetry\label{sec_translational}}
\begin{figure}[h]
\begin{center}
\includegraphics[width=0.8\textwidth]{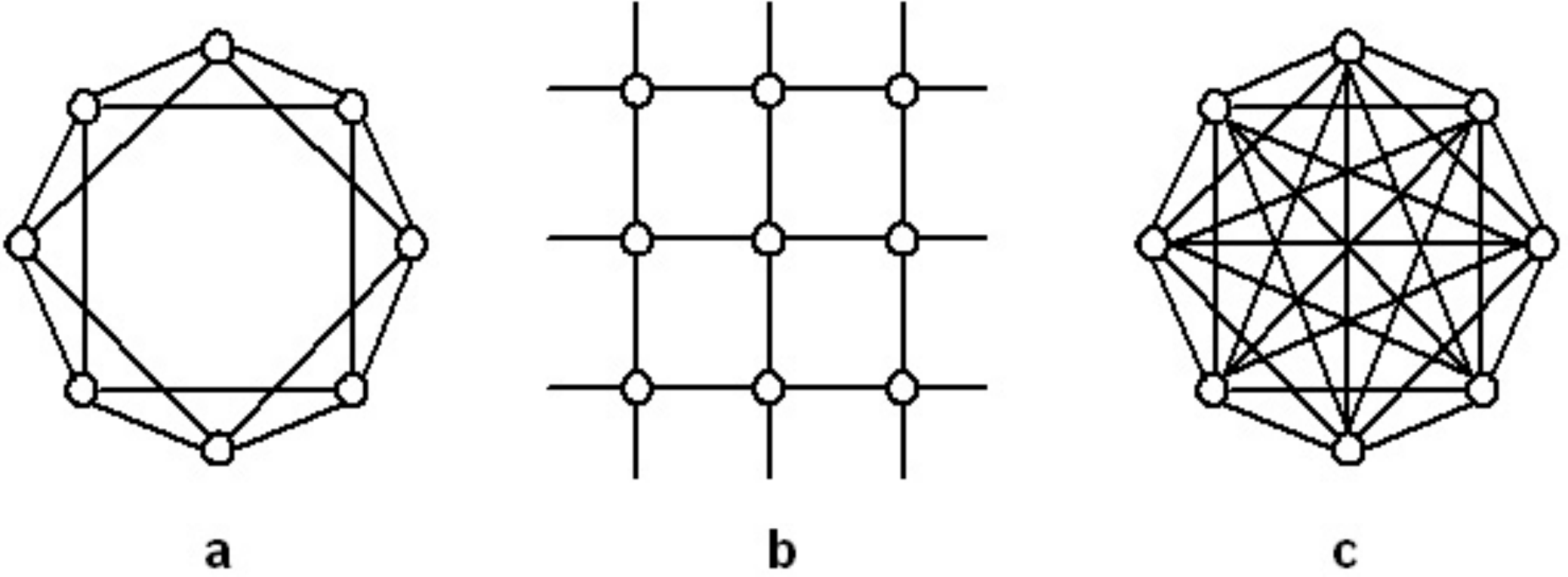}
\end{center}
\caption{Examples of regular graphs. (a) A periodic one-dimensional
($D=1$)  lattice ({\em i.e.} a ring) where each vertex is connected
to its nearest and second-nearest neighbours. (b) A two-dimensional
($D=2$) lattice (in principle of infinite size) where each vertex is
connected only to its nearest neighbours. (c) A complete graph where
every vertex is connected to all other vertices. A complete graph
can be regarded either as a lattice embedded in some space of finite
dimension $D<\infty$ (as in the two previous examples) with
infinite-ranged connections, or as a lattice embedded in infinite
dimension $D=\infty$ with finite-ranged (as in the two previous
examples) connections. \label{fig_regular}} \vspace{-0.3cm}
\end{figure}

As we mentioned, some graphs may be embedded in a metric space where
vertices are assigned positions. In this case, the symmetries
(automorphisms) of a graph are induced by the transformations of
coordinates in the embedding space, even if topologically their are
simply permutations of vertices. This means that the topological
properties of the graph, which are independent of the embedding
space, will nonetheless reflect the properties of the latter. For
instance, lattices are naturally formed by connecting vertices to
their nearest neighbours in some embedding space (see Figure
\ref{fig_embed}). A simple type of discrete symmetry encountered in
(either infinite or periodic) regular lattices is translational
symmetry. That is, the fact that the topology of a lattice embedded
in some $D$-dimensional space does not change after a displacement
by an integer multiple of the lattice spacing. Lattices are a
particular type of \emph{regular graphs}, {\em i.e.} graphs where
every vertex has the same number of neighbours. In Figure
\ref{fig_regular} we show three examples of regular graphs embedded
in different dimensions ($D=1$, $D=2$ and $D=\infty$) and with
differently ranged connections (nearest neighbours, nearest and
second-nearest neighbours, infinite neighbours).

If the labeling of vertices reflects their position in space, then
translational symmetry is reflected in some regularities of the
adjacency matrix $A$ of the network (for undirected graphs, where no
orientation is defined on the edges, the adjacency matrix $A$ is a
binary matrix whose entries equal $a_{ij}=1$ if a link between
vertex $i$ and vertex $j$ is present, and $a_{ij}=0$ otherwise; here
$i=1,\dots, N$ where $N$ is the total number of vertices, {\em i.e.}
the size of the network). For instance, if the vertices are numbered
cyclically along the ring, the adjacency matrices $A_a$ and $A_c$ of
the graphs shown in Figure \ref{fig_regular}a and c read
\begin{equation}\label{eq_mat} A_a=\left(\begin{array}{cccccccc}
0&1&1&0&0&0&1&1\\
1&0&1&1&0&0&0&1\\
1&1&0&1&1&0&0&0\\
0&1&1&0&1&1&0&1\\
0&0&1&1&0&1&1&0\\
0&0&0&1&1&0&1&1\\
1&0&0&0&1&1&0&1\\
1&1&0&1&0&1&1&0
\end{array}\right)\qquad
A_c=\left(\begin{array}{cccccccc}
0&1&1&1&1&1&1&1\\
1&0&1&1&1&1&1&1\\
1&1&0&1&1&1&1&1\\
1&1&1&0&1&1&1&1\\
1&1&1&1&0&1&1&1\\
1&1&1&1&1&0&1&1\\
1&1&1&1&1&1&0&1\\
1&1&1&1&1&1&1&0
\end{array}\right)
\vspace{0.3cm}
\end{equation}
respectively. Translational symmetry is one of the traditional
assumptions used in the theoretical study of discrete (or
discretized) dynamical systems, and most of the available analytical
results about dynamical processes are only valid under the
assumption of the existence of this symmetry.

However, as one moves beyond the simple case of atoms regularly
embedded in crystal lattices, virtually all real-world networks
strongly violate translational symmetry. An important deviation from
lattice-like topology in real networks is signaled by a surprisingly
small value of the average \emph{inter-vertex distance}, {\em i.e.}
the average number of links one needs to traverse along the shortest
path connecting two vertices. In most real networks, this quantity
increases at most logarithmically with the number  $N$ of vertices,
a phenomenon known as the \emph{small-world} effect
\cite{guidosbook}. This behavior is also encountered in the random
graph model mentioned in Section \ref{sec_levels} but not in
lattices, where the average distance (if infinite-ranged connections
are not allowed, e.g. for the graphs in Figure \ref{fig_regular}a
and b but not for that in Figure \ref{fig_regular}c) grows as
$N^{1/D}$, thus much faster. The breakdown of translational symmetry
implies that the wealth of knowledge accumulated in the literature
about the outcome of dynamical processes on lattices cannot be
applied to the same processes when they take place on real networks
\cite{dynamicalprocessesoncomplexnetworks}. We already mentioned
epidemic spreading processes as an example of the surprising
deviation between dynamics on lattices and on more complicated
networks. Nonetheless, real networks bear an interesting similarity
with regular graphs, namely a large average value of the
\emph{clustering coefficient}, defined as the number of triangles
(loops of length three) starting at a vertex, divided by its maximum
possible value.

The simultaneous presence of a small average distance and of a large
clustering coefficient (which is sometimes taken as a stronger
definition of the \emph{small-world} effect) has motivated the
introduction of an important and popular network model which is
somehow `intermediate' between regular lattices and random graphs.
In the model proposed by Watts and Strogatz \cite{smallworld}, one
starts with a regular lattice and then, with fixed probability $p$,
goes through every edge and rewires one of its two end-point
connections to a new, randomly chosen vertex. Clearly, when $p=0$
one has the original lattice (large clustering and large distance),
while when $p=1$ one has a completely random graph (small clustering
and small distance). Thus the parameter $p$ can be viewed as a
measure of the deviation from complete translational symmetry in the
model. Interestingly, in a broad intermediate range of values one
simultaneously obtains a large clustering and a small distance, thus
recovering the empirically observed effect. This suggests that real
networks may be partially, but surely not completely, affected by
translational symmetry (due for instance to the existence of a
natural spatial embedding). As we shall discuss in Section
\ref{sec_properties}, translational symmetry, and in general the
dependence of structural properties on the vertices' positions in
some embedding space, is an example of a more general situation
where vertices are characterised by some non-topological quantity
that may determine or condition their connectivity patterns.

\subsection{Scale Invariance\label{sec_scaleinvariance}}
\begin{figure}[h]
\begin{center}
\includegraphics[width=0.4\textwidth]{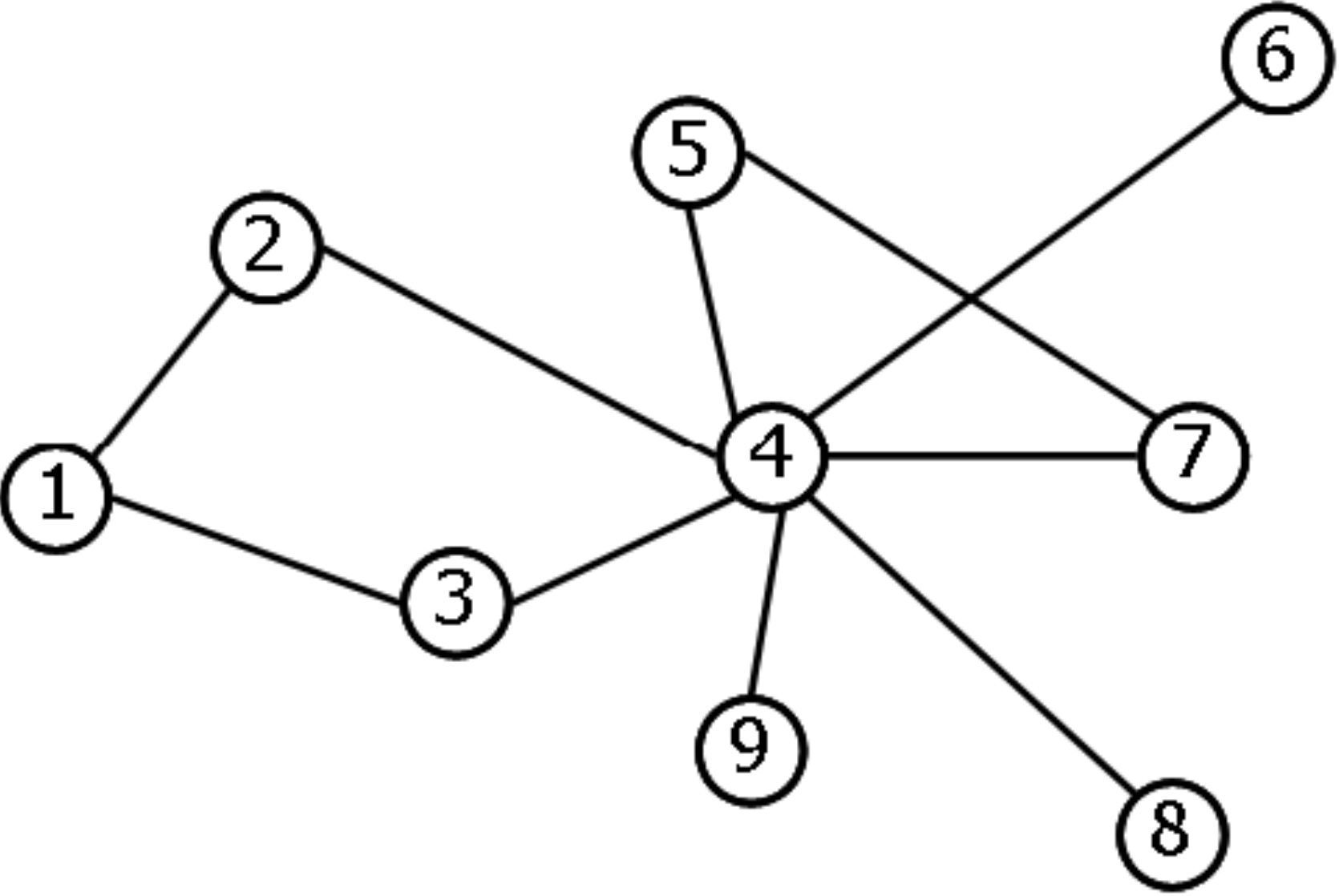}
\end{center}
\caption{Example of a network with $N=9$ vertices and highly
heterogeneous degree distribution. Vertex $4$ is a highly connected
hub with degree $k_4=7$ (the maximum possible value is $N-1=8$),
whereas all other vertices have only $k=1$ or $k=2$ connections.
\label{fig_scalefree}} \vspace{-0.3cm}
\end{figure}

As we mentioned, one of the most striking and ubiquitous features of
real networks is the power-law form $P(k)\propto k^{-\gamma}$ of the
degree distribution. This property means that vertices are extremely
heterogeneous in terms of their number of connections: many vertices
have a few links, and a few vertices (the \emph{hubs}) have
incredibly many links. An example of a small network with highly
heterogeneous degree distribution is shown in Figure
\ref{fig_scalefree}. Importantly, most of the empirically observed
values of the exponent $\gamma$ are found to be in the range
$2<\gamma<3$, where the variance of the distribution diverges. This
implies that there is no typical scale for the degree $k$ in the
system, and motivates the expression \emph{scale-free \linebreak
network} \cite{guidosbook}.

The above property is an example of a remarkable type of symmetry,
precisely scale invariance. It is found across different domains
\cite{powerlaws}, and in particular in fractal objects. In fractals,
scale invariance is manifest in the fact that iterated
magnifications of an object all have the same shape, {\em i.e.} the
system `looks the same' at all scales. Similarly, in networks one
finds that if the scale of the observation is changed (e.g. one
switches from degree $k$ to degree $ak$, with $a$ positive), the
number of vertices with given degree only changes by a
(magnification) factor, from $P(k)$ to $P(ak)=a^{-\gamma} P(k)$.
This is very different from exponential distributions, characterized
by a strong variation in the number of counts as the scale is
changed. In networks, power laws have also been found to describe
the distribution of link weights, of the sum of link weights (the
so-called strength) of vertices, and of many more \linebreak
quantities \cite{guidosbook}. They also appear to hold across
various coarse-grained levels of description of the same network, if
groups of vertices are iteratively merged into `supervertices' and
the original connections collapsed into links among these
supervertices \cite{shlomo}. The symmetry group associated to scale
invariance, {\em i.e.} the \emph{renormalization group}
\cite{renormalization}, has therefore been used many times to
theoretically understand power-law distributed network properties.

The presence of a scale-free topology across several real-world
networks, which is not reproduced by the Erd\H{o}s-R\'enyi model and
by the Watts-Strogatz one, has led to the introduction of new
theoretical mechanisms that could possibly explain the onset of this
widespread phenomenon. The earliest (even if analogous mechanisms
were already known in different contexts \cite{powerlaws}) and most
popular scale-free network model is the one proposed by Barabasi and
Albert \cite{BA}. It is based on two key ideas: firstly, networks
can grow in time, therefore one can assume that new vertices are
continuously added to a preexisting network; secondly, already
popular (highly connected) vertices are likely to become more and
more popular (`rich get richer'). The latter idea, known as
\emph{preferential attachment}, is modeled as a multiplicative
process in degree space: the probability that newly introduced
vertices establish a connection to a preexisting vertex $i$ is
proportional to the degree $k_i$ of that vertex. The iteration of
this elementary process of growth and preferential attachment
eventually generates a power-law degree distribution of the form
$P(k)\propto k^{-3}$. In degree space, preferential attachment is a
symmetry-breaking mechanism: vertices are not equally likely to
receive new connections as the network grows. Even if all vertices
are identical \emph{a priori}, preferential attachment determines
and amplifies heterogeneities in the degree, and eventually vertices
with different degrees become subject to different probabilistic
rules. Since in the model there is a tight relationship between the
degree of a vertex and the time the same vertex entered the network,
one could also say that different injection times imply different
expected topological properties. On the other hand, with respect to
scale invariance, preferential attachment is symmetry-preserving and
gives rise to a stationary process. Indeed, as the network grows
infinitely in size over time, its scale-free degree distribution
remains unchanged. This highlights how the same network properties
may bear different meanings in relation to different symmetries.
There are now many alternative models that reproduce scale-free
networks with any value of the power-law exponent $\gamma$, not only
$\gamma=3$ \cite{guidosbook,largescalestructure,adaptivenetworks}.
In all of them, there is some mechanism that eventually sets on and
drives the network to converge to an extremely heterogeneous
topology. We shall describe one of these \linebreak models
\cite{fitness} in Section \ref{sec_properties}. Before doing that,
in the following Sections \ref{sec_structural} and
\ref{sec_statistical} we shall make a more general discussion about
symmetry breaking due to differences in topological properties in a
model-free and real-world framework.

\subsection{Graph Automorphisms and Structural Equivalence\label{sec_structural}}
Various types of vertex permutation symmetry can be defined for
graphs. Some of these symmetries are trivial, while others can be
very interesting and informative. A trivial example is the symmetry
under any overall permutation of vertex labels: if all vertices are
relabelled differently, and a new adjacency matrix is defined
accordingly, the resulting graph will have exactly the same topology
of the original one ({\em i.e.} the two graphs are \emph{isomorphic}
to each other \cite{harary}). Since one is always free to assign any
labelling to vertices, permutation symmetry trivially holds in any
network (in mathematical words, an unlabelled graph is invariant
under vertex relabelling). In this sense, a graph with $N$ vertices
is trivially invariant under the possible $N!$ permutations of
vertex labels, if all edges are relabelled accordingly.

\begin{figure}[h]
\begin{center}
\includegraphics[width=0.4\textwidth]{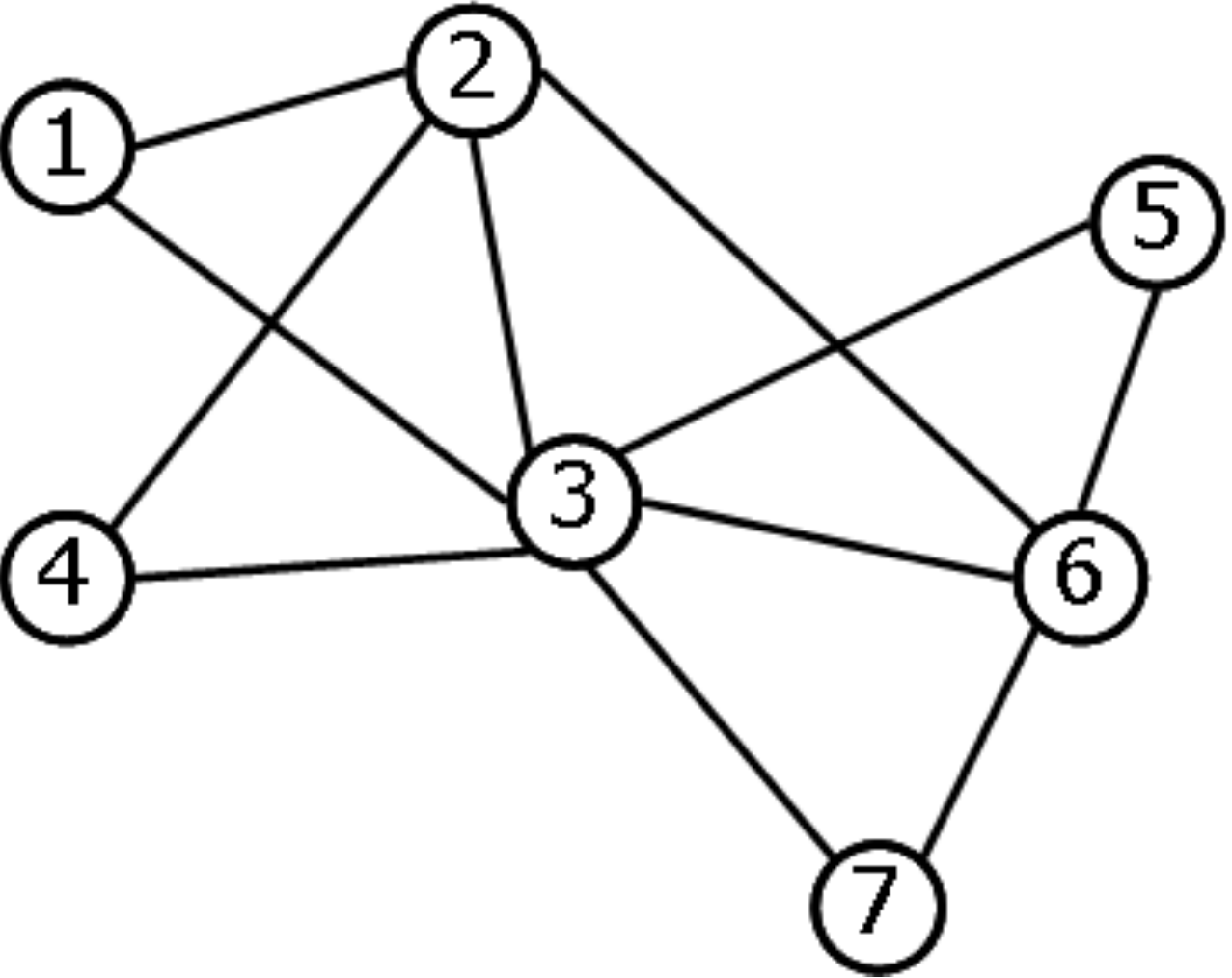}
\end{center}
\caption{In the example shown, vertices $1$ and $4$ are structurally
equivalent because they have the same set of neighbours (vertices
$2$ and $3$). Similarly,  vertices $5$ and $7$ are structurally
equivalent because they are both connected only to vertices $3$ and
$6$. \label{fig_equivalence}} \vspace{-0.3cm}
\end{figure}

However, a far less trivial problem is whether, after a given
labelling has been chosen (and the graph has therefore become a
labelled one), the network still remains invariant under further
vertex permutations. As we mentioned in Section
\ref{sec_transformations}, this is the \emph{graph automorphism}
problem, {\em i.e.} the analysis of the isomorphisms of a graph with
itself \cite{harary}. Suppose the identity of every vertex has been
fixed by assigning a unique label to each of them (as we mentioned,
this labelling is arbitrary and every choice leads to an equivalent
description of the same network). Once a labelling is chosen, one
may still find that a particular graph is unchanged after
permutations of some vertices (without exchanging the identity of
the latter). Graph automorphisms are studied in detail by discrete
mathematics. Technically, the set of vertex permutations defining
the automorphisms of a graph forms a symmetry group, denoted as the
\emph{automorphism group} of the graph. Given a particular graph,
the analysis of its automorphism groups provides a characterization
of its properties, and in particular its symmetries. Traditionally,
automorphism groups are studied for specific classes of graphs
generated according to deterministic rules, which represent standard
examples in graph theory \cite{harary}. The analysis of the
automorphism groups of real-world networks is instead very recent
\cite{symmetry,quotient,redundancy,symmetry_wtw}. One of the reasons
why it is interesting to look for automorphisms in real networks is
their relation to the following important problem. If two vertices
$i$ and $j$ have exactly the same set of neighbors (irrespective of
whether they are neighbors of each other), then a permutation
exchanging $i$ and $j$, and leaving all other vertices unchanged,
leads to exactly the same graph. In social science, when this occurs
the vertices $i$ and $j$ are said to be \emph{structurally
\linebreak equivalent} \cite{wasserman}. In food web ecology (where
also the direction of each link to the common neighbours must be the
same), they are said to belong to the same \emph{trophic species}
\cite{ecologicalnetworks,myfoodwebs}. An illustration of structural
equivalence is shown in Figure \ref{fig_equivalence}. The adjacency
matrix of a graph where $i$ and $j$ are structurally equivalent is
unchanged after exchanging its $i$th and $j$th row, and its $i$th
and $j$th column. In doing so, we are not interchanging the identity
of $i$ and $j$, which still represent the original vertices (for
instance, two particular persons in a social network).

Structural equivalence, which may or may not be present in a given
real network, is very important for many disciplines. It is directly
related to the problem of network robustness: if a vertex is removed
from the network, the presence of at least one structurally
equivalent vertex warrants that there are no secondary effects
(other vertices becoming disconnected) or major topological changes.
By contrast, the effects can be dramatic if the removed vertex is a
special one with no structurally equivalent peers (for instance, a
highly connected hub). The analysis of the automorphism groups of
real networks has revealed that, unlike random graphs, real networks
are highly symmetric and contain a significant amount of structural
redundancy \cite{symmetry,quotient,redundancy,symmetry_wtw}. This
property may naturally arise from growth processes involved in the
formation of many  networks, and affects local topological
properties such as network motifs (subgraphs of three or four
vertices recurring in real networks much more often than in random
\linebreak graphs \cite{motifs}). Graph automorphisms have also been
used to simplify the topology of real networks by collapsing
redundant information and obtaining \emph{network quotients}
\cite{quotient}, {\em i.e.} coarse grained graphs without structural
repetition. Despite quotients of real networks are substantially
smaller than the original graphs, they are found to preserve various
structural properties (degree heterogeneity, small distance, {\em
etc.}), effectively capturing a sort of skeleton of the entire
empirical networks \cite{quotient,symmetry_wtw}.

\subsection{Statistical Equivalence\label{sec_statistical}}
Structural equivalence is a very strict definition of similarity
between two vertices. A more relaxed condition that is usually of
interest in sufficiently large networks is whether two vertices are
\emph{statistically equivalent}, {\em i.e.} whether their
topological properties are the same in an average or weak sense. For
instance, one could ask whether two vertices $i$ and $j$ have simply
the same degree (irrespective of the identity of their neighbors),
and/or the same number of second neighbours, or whether they
participate in the same number of triangles and/or longer loops.
Similarly, one could be interested in finding two vertices whose
neighbours have the same average degree, irrespective of the numbers
of neighbours of each vertex, and of the individual values of the
degrees of these neighbours (this is explained in more detail
below). In all these examples, one focuses on a subset (or some
average value) of the possible topological properties involving $i$
and $j$, and defines an equivalence with respect to it only.
According to this relaxed condition, a number of statistically
equivalent vertices are found in real networks. The structure of the
resulting equivalence classes determines the symmetry of a
particular network. While permutations of structurally equivalent
vertices are exact symmetries of the graph ({\em i.e.}
automorphisms), permutations of statistically equivalent vertices
are stochastic symmetries in the sense introduced in Section
\ref{sec_transformations}. Such transformations do not map a network
to itself, but to another member of the family of networks with the
same statistical properties. Importantly, while even small errors
such as a missing link in the data have a dramatic effect on
structural equivalence, statistical equivalence is more robust to
fluctuations in network structure. Moreover, introducing this
stochastic type of symmetry gives rise to identify more general
patterns than those accessible to the analysis of structural
equivalence. We discuss this concept by making some examples of the
main scientific questions related to statistical equivalence in
networks.

\emph{Do all vertices in a network have the same degree?} As already
discussed in Section \ref{sec_scaleinvariance}, this type of
symmetry is strongly violated in real networks. A weaker question
would be: are the degrees of all vertices \emph{nearly} the same? In
this case, one could speak of a typical degree of vertices, and
interpret the deviations from the average value as finite
fluctuations due either to external noise or some intrinsic
stochasticity. However, as we mentioned, the majority of real
networks are scale-free, with degrees being broadly distributed and
wildly fluctuating. There are many vertices with small degree, among
which one can in principle find vertices with exactly the same
number of neighbors, but also a few vertices with extremely large
degree, which strongly break the symmetry.

\emph{Is the average degree of the neighbors of all vertices
(nearly) the same?} After recognizing that some vertices attract
many more links than others, one can move one step forward and
wonder what is the average degree of the neighbors of a given vertex
(the so-called \emph{average nearest neighbor degree}, or ANND
\cite{guidosbook}). This quantity encodes some information about the
matching patterns in the network: if the degree plays no role in
deciding whether two vertices are connected, then one expects that
the ANND is independent of the degree itself (as we discuss below,
this is not completely true). By contrast, one finds the presence of
strong correlations between the degrees of neighboring vertices.
These correlations can be either positive or negative, and have
opposite effects on the ANND. In networks where large-degree
vertices are more likely to be connected to each other than to
low-degree ones, one observes an increasing trend of the ANND as a
function of the degree. This property is known as
\emph{assortativity} \cite{newman_assortative}. In networks where
the opposite is true, the ANND decreases with the degree, a
situation denoted \emph{disassortativity}. Importantly,
degree-degree  correlations have profound effects on the outcomes of
dynamical processes taking place on networks
\cite{dynamicalprocessesoncomplexnetworks}.

\emph{Do all vertices have (nearly) the same clustering
coefficient?} Again, this symmetry is generally not observed, as
vertices with different degree also have different values of the
clustering coefficient. The latter usually displays a decreasing
trend with the degree $k$. This behavior has been interpreted as the
signature of a hierarchically organised topology, where a simple
wiring pattern is repeated at different scales in a bottom-up
fashion: first creating modules of vertices, then modules of
modules, {\em etc.} \cite{hierarchy}. Since both the clustering
coefficient and the ANND strongly depend on the degree, and since
the latter is broadly distributed, it appears that real networks are
characterised by a high level of complexity, with no characteristic
scale associated to any of the simplest topological properties one
can define.

However, the last observation also leads to a reverse, possibly
simplifying, approach to the problem. Interestingly, it has been
shown that some of the correlations mentioned above are partly an
unavoidable, `spurious' outcome of enforcing some topological
constraints in the network \cite{maslov,newman_origin}. That is,
exactly because many properties ultimately depend on the degree, a
number of structural patterns are automatically generated once the
degrees of all vertices are fixed to specified values. For instance,
in networks with power-law degree distribution the ANND and the
clustering coefficient both decrease with the degree. These patterns
do not signal `true' higher-order correlations, as they are natural
outcomes due to the presence of simpler constraints.  If an
explanation from the latter exists, it also automatically explains
the former. This  highlights the importance of separating low-order
effects from more fundamental higher-order structural patterns. This
problem leads to the definition of suitable \emph{null models} of
networks, a point that we shall discuss in Section
\ref{sec_equiprobability}.

\subsection{Invariance under Permutation of External Properties\label{sec_properties}}
\begin{figure}[h]
\begin{center}
\includegraphics[width=0.8\textwidth]{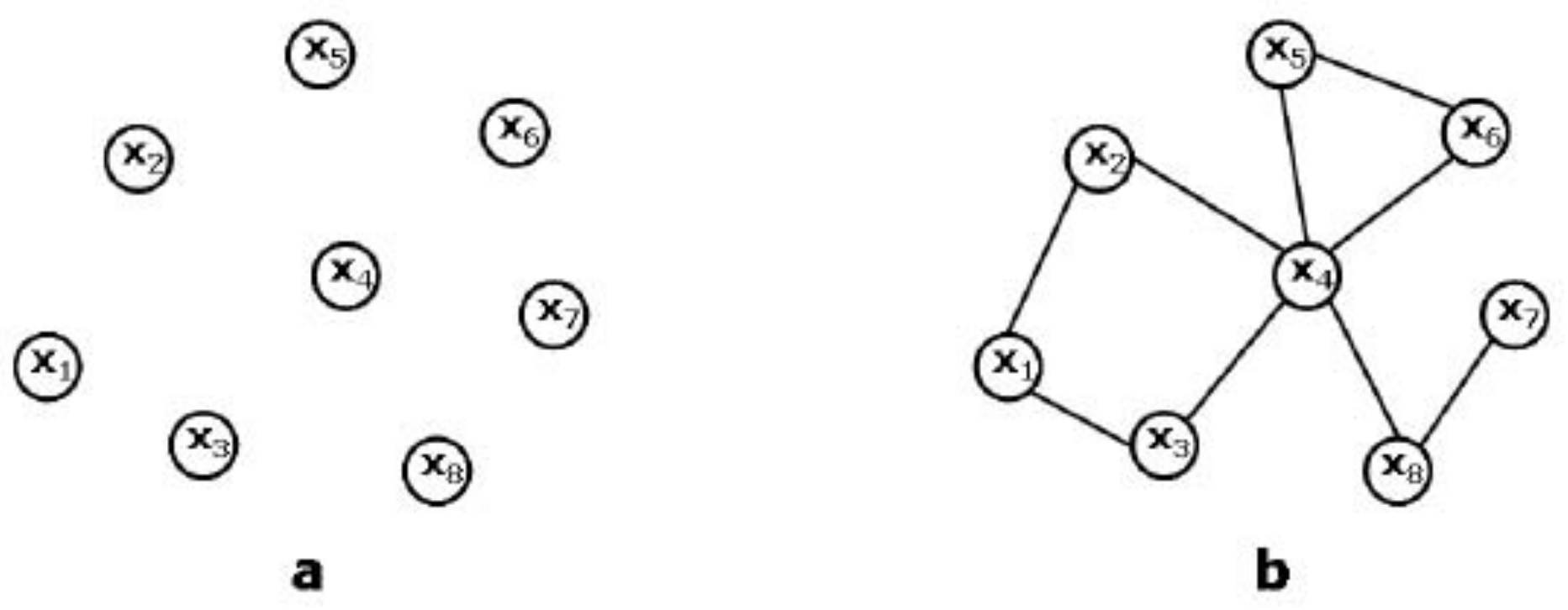}
\end{center}
\caption{The topological properties of a network may depend on some
external property $x$ attached to vertices. (a) For instance, in the
\emph{fitness} model \cite{fitness} one starts with an empty network
where each vertex $i$ is assigned a fitness value $x_i$ drawn from
some specified distribution $\rho(x)$. (b) Then, a link between
vertices $i$ and $j$ is drawn with probability $p(x_i,x_j)$.
Vertices with identical values of $x$ are statistically equivalent:
all their topological properties have the same expected values.
However, the probabilistic nature of the model implies that, in a
particular realization of the network, two vertices $i$ and $j$ with
$x_i=x_j$ are not necessarily structurally equivalent, and
conversely two structurally equivalent vertices (for instance, $x_2$
and $x_3$ in the example shown) do not necessarily have identical
fitness values (as we may have $x_2\ne x_3$; indeed, this is
typically the case if $x$ is drawn from a continuous probability
density). This highlights the difference between structural
equivalence and statistical equivalence. \label{fig_fitness}}
\vspace{-0.6cm}
\end{figure}

An important type of permutation symmetry can be defined when some
external, non-topological property is attached to vertices (or to
edges, or to other subgraphs; but we will consider the case of
vertices for simplicity). This situation is particularly relevant
when one is interested in studying the relation between the topology
and some other property characterising the vertices of a network,
and is tightly related (even if in a nontrivial way) to structural
and statistical equivalence, as the example in Figure
\ref{fig_fitness} shows. Note that translational symmetry (described
in Section \ref{sec_translational}) can be viewed as a particular
case of this problem, if vertices are assigned positions in some
metric space. Translational symmetry is in principle an exact
symmetry (the graph is mapped onto itself) since it is the effect of
a deterministic graph formation rule. However, symmetries due to
external properties are in general stochastic in the sense discussed
in Section \ref{sec_transformations}, since real networks are always
best understood as a result of non-deterministic rules. We therefore
expect that stochastic symmetry is more powerful in detecting
patterns in real networks than exact symmetry, and the following
discussion confirms this expectation.

The impact of external factors is an extremely important problem,
related to key questions about network formation, for many research
areas. Typical examples include: \emph{is a social network partly
determined by factors such as race, gender, age, etc.?} \emph{Is
wealth or income relevant to the formation of economic networks?} In
order to answer the above questions, one needs a way to assess the
structural impact of properties which are in some sense external to
the network.

There have been many attempts in this direction. Social network
analysis has a long tradition in dealing with this problem, firmly
based on statistical theory. The role of vertex properties is
generally inspected through the values of regression parameters used
in suitable graph models that are fitted to the real network
\cite{wasserman}. More recently, in the physics community different
approaches have been proposed. Techniques have been introduced
\cite{newman_assortative,pin_jackson} in order to capture whether
the connections observed in a particular network occur mainly
between vertices with similar properties (this is a generalised
notion of \emph{assortativity}, not necessarily related to vertices'
degrees, also known as \emph{homophily} in social science) or
between vertices with different properties (\emph{disassortativity})
. More generally, there have been attempts in understanding whether
a specification of vertex properties effectively reduces the
available configuration space for a real network \cite{pin_ginestra}
and can thus be interpreted as a  structurally important factor. All
these different approaches to the same problem could be restated in
more general terms as follows: \emph{is the network (stochastically)
symmetric under a permutation of the properties attached to
vertices?} If this is the case, the properties under consideration
have no statistically significant impact on network structure.
Otherwise, vertex-specific features are symmetry-breaking, as
vertices with different properties are no longer equivalent under a
somewhat generalised notion of the statistical equivalence described
in \linebreak Section \ref{sec_statistical}. In particular, the
overall permutation symmetry of vertex properties is broken and the
network is only symmetric under a restricted set of permutations
exchanging vertices within the same equivalence classes (sets of
vertices with the same external properties). It is therefore clear
that the behaviour of a network under the permutations associated to
this type of permutation symmetry is determined by, and carries
information about, the effects that external quantities have on the
topology.

In general, the behaviour of a real network under permutation of
external properties can be very complicated and lead to a variety of
different symmetry properties. However, it is possible to understand
the problem clearly in simplified models. Indeed, the idea that
vertex properties may be crucial to network formation has led to the
definition of an important class of network models known as
\emph{fitness} or \emph{hidden variable} models \cite{fitness}.
Unlike the Barabasi-Albert model mentioned in Section
\ref{sec_scaleinvariance}, fitness models are static and do not
require the hypothesis of network growth. In these models, one
assumes that the probability $p_{ij}$ that a link is present between
vertex $i$ and vertex $j$ is a function $p(x_i,x_j)$ of some
property $x$, or \emph{fitness}, attached to these vertices (see
Figure \ref{fig_fitness}). Therefore the model requires the
specification of a list of fitness values $\{x_i\}$, usually assumed
to be drawn independently from some probability distribution
$\rho(x)$, and of the connection function $p(x_i,x_j)$. All the
expected topological properties crucially depend on $\{x_i\}$. For
instance, the expected degree of two vertices $i$ and $j$ with
different fitness values ($x_i\ne x_j$) is in general different. On
the other hand, two vertices with $x_i= x_j$ are statistically
equivalent. However, due to the probabilistic nature of the model,
in a particular realization of the network the statistical
equivalence of vertices with equal fitness values does not
necessarily reflects in their structural equivalence (see example in
Figure \ref{fig_fitness}). This model specification successfully
reproduces the situation mentioned above, as the permutation
symmetry of vertex properties is broken down to disjoint equivalence
classes represented by sets of vertices with identical hidden
values. Moreover, the flexibility in the choice of the fitness
values and  connection probability allows to reproduce various
topological properties of real-world networks. For instance, a
power-law distribution of fitness values (mimicking some
heterogeneously distributed real-world feature such as individual
wealth, country population, {\em etc.}) and a connection probability
that linearly depends on the fitness naturally lead to a scale-free
network topology \cite{fitness}. Besides providing a valid route to
network modelling, hidden variable models can also be fitted to real
networks and shed light on the presence of external factors case by
\linebreak case \cite{mylikelihood,ramasco}. In particular, inverse
methods have been devised in order to extract, only from the
topology of a real network, the values of the hidden variables
$\{x_i\}$ potentially related to network formation. These values can
then be compared with the values of candidate external properties
relevant to that particular network, a strategy that has been shown
to successfully identify key factors related to structure in
real-world cases \cite{mylikelihood}.

\subsection{Ensemble Equiprobability\label{sec_equiprobability}}
As we anticipated in Section \ref{sec_transformations}, there are
important symmetries associated not to a single graph, but to a
\emph{statistical ensemble} of graphs (we will define a graph
ensemble rigorously below). If the ensemble is a good model of a
real network, these symmetries can then be naturally related to the
real network itself. This possibility allows us to illustrate in
more detail our idea of stochastically symmetric ensemble, and the
definition of stochastically symmetric graph as a network which is
well reproduced by a stochastically symmetric ensemble (see Section
\ref{sec_transformations}). Null models automatically come into play
when one is interested in understanding whether, in a given network,
complicated high-order topological properties can be traced back to
simpler low-level constraints. We already mentioned this problem in
Section \ref{sec_statistical}. In order to answer this question, it
is necessary to consider a null model by generating a collection of
graphs having some property in common with the real network (these
properties act therefore as constraints), and being completely
random otherwise. This amounts to generate an ensemble of graphs
that maximizes an \emph{entropy}, that we shall define in a moment,
under the enforced constraints. Then, one can compare the properties
of the real network with the corresponding averages over the
randomised ensemble. If there is no statistically significant
difference, one can conclude that the constraints considered are
indeed enough in order to generate all the other properties of the
real network. If differences are significant, then there are other
factors shaping the observed topology. We now rephrase this idea
more formally, and show how it highlights an intimate and
instructive connection between symmetry, entropy and complexity
\linebreak in networks.

A statistical ensemble of graphs \cite{newman_statistical} is a
collection of $M$ graphs $\{G_1,G_2,\dots, G_M\}$, each with an
associated occurrence probability $P(G)$ satisfying
\begin{equation}
\sum_G P(G)\equiv \sum_{m=1}^M P(G_m)=1
\end{equation}
We already mentioned examples of graph ensembles, without explicitly
noticing it: Erd\H{o}s-R\'enyi model (Sections \ref{sec_levels} and
\ref{sec_translational}), the Watts-Strogatz model (Section
\ref{sec_translational}), the Barabasi-Albert model (Section
\ref{sec_scaleinvariance}) and the fitness model
(\ref{sec_properties}) are all examples of collections of possible
graphs generated by probabilistic rules. The Barabasi-Albert model
is a non-equilibrium ensemble, as it generates networks growing
indefinitely in time; all the other examples mentioned above are
instead equilibrium ensembles. In what follows, we restrict
ourselves to the equilibrium case. Each graph $G$ is uniquely
specified by its adjacency (or weight) matrix, so we can think of
$G$ as of a matrix. For instance, if one is interested in the
ensemble of binary undirected graphs with $N$ vertices and no
self-loops (edges starting and ending at the same vertex), then $G$
will be a symmetric Boolean matrix with zeroes along the diagonal,
and there will be $M=2^{N(N-1)/2}$ possible such matrices in the
ensemble. In order to generate a maximally random ensemble of graphs
with given constraints
\cite{newman_statistical,ginestra_entropy,mybosefermi}, one needs to
find the form of the probability $P(G)$ that maximises the
Shannon-Gibbs entropy
\begin{equation}
S\equiv -\sum_G P(G)\ln P(G)
\label{eq_entropy}
\end{equation}
(a standard measure of disorder or uncertainty) under the enforced
constraints. The latter are a collection $\{c_1,\dots, c_K\}$ of $K$
topological properties, forming a $K$-dimensional vector $\vec{c}$.
Each property \linebreak $c_a$ ($a=1,\dots, K$) evaluates to
$c_a(G)$ when measured on the particular graph $G$. If the ensemble
is meant as a null model of an empirical network $G^*$, the
constraints will be chosen as the properties $\vec{c}(G^*)$
evaluated on the particular graph $G^*$.

There are various possible choices to solve the entropy maximisation
problem, and different ensembles that one can define accordingly. If
one is interested in matching the constraints \emph{exactly}, {\em
i.e.} in picking out only the graphs that have exactly the same
properties as a given network $G^*$, then the solution is given by
the probability
\begin{equation}
P(G)=\left\{\begin{array}{ll}1/\mathcal{N}[\vec{c}(G^*)]&\textrm{if } \vec{c}(G)=\vec{c}(G^*)\\
0&\textrm{otherwise}
\end{array}\right.
\end{equation}
where $\mathcal{N}[\vec{c}(G^*)]$ is the number of graphs matching
the constraints $\vec{c}(G^*)$. The above probability is uniform
over the set of configurations matching the constraints exactly, and
the resulting ensemble is known in statistical physics as the
\emph{microcanonical} ensemble. With the above choice, the entropy
defined in Equation (\ref{eq_entropy}) takes the form
\begin{equation}
S= -\mathcal{N}[\vec{c}(G^*)] \frac{1}{\mathcal{N}[\vec{c}(G^*)]}\ln \frac{1}{\mathcal{N}[\vec{c}(G^*)]}=\ln \mathcal{N}[\vec{c}(G^*)]
\label{eq_microentropy}
\end{equation}
which is known as the \emph{microcanonical entropy} and is simply
the logarithm of the number of configurations exactly matching the
constraints.

A second alternative consists in requiring that the constraints
$\vec{c}$ are matched \emph{on average}, {\em i.e.} allowing any
graph to occur with non-zero probability, provided that the expected
value $\langle \vec{c}\rangle=\sum_G P(G)\vec{c}(G)$ of the
constraints matches the required value $\vec{c}(G^*)$. This problem
can be solved introducing Lagrange multipliers $\{\theta_1,\dots,
\theta_K\}$, each associated to one of the constraints. The solution
is the probability distribution \be P(G)=\frac{e^{-H(G)}}{Z} \ee
where $H(G)$ (the \emph{graph Hamiltonian}) is a linear combination
of the constraints \be H(G)\equiv \sum_{a=1}^K \theta_a c_a(G) \ee
and $Z$ is the \emph{partition function} that properly normalizes
the probability: \be Z\equiv\sum_G e^{-H(G)} \ee Thus both $Z$ and
$P(G)$ depend on the $K$ parameters $\{\theta_1,\dots, \theta_K\}$.
The ensemble generated by the above probability is known in physics
as the \emph{canonical} ensemble. For a given choice of the
parameters $\{\theta_1,\dots, \theta_K\}$, the expected value of a
topological property $X$ across the ensemble is
\begin{equation}
\langle X(\theta_1,\dots, \theta_K)\rangle\equiv\sum_G P(G) X(G)
\label{eq_X}
\end{equation}
(throughout this review, the angular brackets $\langle \cdot\rangle$
will denote ensemble averages). In order to match the constraints
$\vec{c}(G^*)$ on average, the $K$ parameters $\{\theta_1,\dots,
\theta_K\}$ must be set to the particular values
$\{\theta^*_1,\dots, \theta^*_K\}$ such that
\begin{equation}
\langle c_a(\theta^*_1,\dots, \theta^*_K)\rangle=c_a(G^*)\qquad a=1,\dots,K
\end{equation}
Importantly, the above parameter choice corresponds with what the
\emph{maximum likelihood principle} would indicate
\cite{mylikelihood}, {\em i.e.} with the values maximising the
probability $P(G^*)$ to obtain the real network $G^*$ under the
model considered. We will indicate the maximum-likelihood parameter
choice explicitly in the examples considered later on. It has been
shown \cite{newman_statistical} that the canonical ensemble of
networks coincides with the \emph{exponential random graph models}
that have been first introduced in social science \cite{wasserman}.
The Hamiltonian $H(G)$ represents the \emph{energy}, or \emph{cost},
associated with a given configuration, and contains all the
information required in order to formally obtain $P(G)$. This means
that any two graphs $G_1$ and $G_2$ for which
\begin{equation}
H(G_1)=H(G_2)
\end{equation}
have the same ensemble probability $P(G_1)=P(G_2)$. Thus, the
symmetries of $H(G)$ are transformations connecting equiprobable
graphs in the ensemble. Such transformations map a graph $G_1$ into
a graph $G_2$ which has a different topology but exactly the same
values of the enforced constraints. According to our definition in
Section \ref{sec_transformations}, a canonical graph ensemble is
stochastically symmetric under such transformations. If a canonical
graph ensemble is a good model of a real network $G^*$, the latter
is also stochastically symmetric. Maximally random graphs with
constraints therefore represent ideal candidates to illustrate the
concept of stochastic symmetry. The symmetries of the Hamiltonian,
together with the parameter values enforcing the constraints,
determine the entropy $S$ of the ensemble. This entropy is a measure
of the residual uncertainty about the detailed topology of a
network, once the constraints are fixed.

In statistical physics, there is also a third class of ensembles,
{\em i.e.} \emph{grandcanonical} ensembles. In the latter, the
number of particles of the system is also allowed to vary, and it is
treated as one of the properties to be matched on average. In the
case of networks, the role of particles is played by \linebreak
links \cite{newman_statistical}, whose number is allowed to vary
already in the canonical ensemble, as the examples considered below
illustrate. Therefore there is no fundamental difference between the
canonical and grandcanonical ensembles of graphs, unless one is
interested in networks with different types of \linebreak links
\cite{mymultispecies}. For large systems, the microcanonical,
canonical and grandcanonical ensembles give very similar results.
The canonical and grandcanonical ensembles have the enormous
advantage to be analytically treatable, as a consequence of the
relaxed requirement on the constraints. For this reason, in what
follows we shall consider (grand)canonical ensembles of graphs.

We now discuss some examples. If we consider again the ensemble of
all possible undirected graphs with $N$ vertices, the completely
symmetric case is the one where each graph $G$ has the same energy
\begin{equation}
H(G)=H_0
\end{equation}
where $H_0$ is a constant. In other words, in this case there are no
constraints. Clearly, each of the $M$ possible graphs has the same
probability
\begin{equation}
P(G)=2^{-N(N-1)/2}
\end{equation}
and therefore the graph probability is uniformly distributed across
the ensemble (in this particular case, the microcanonical and
canonical ensembles coincide). Transformations changing a graph $G$
into any other graph in the ensemble are symmetries of the
Hamiltonian, and lead to the same ensemble probability. Thus this
ensemble is stochastically symmetric under any transformation. The
entropy is the maximum possible, and its value is
\begin{equation}
S=\frac{N(N-1)}{2}\ln 2
\label{eq_maxent}
\end{equation}

A different case is when there is a constraint on the total number
of links $L=\sum_{i<j}a_{ij}$. Then
\begin{equation}
H(G)=\theta L(G)
\label{eq_Hrand}
\end{equation}
and it can be easily shown that
\begin{equation}
P(G)=p^{L(G)}(1-p)^{N(N-1)/2-L(G)}
\label{eq_prandom}
\end{equation}
where $p\equiv e^{-\theta}/(1+e^{-\theta})$. This shows that, as
expected, two graphs $G_1$ and $G_2$ with the same number of links
$L(G_1)=L(G_2)$ are equiprobable. Graph transformations preserving
this number are symmetries of the Hamiltonian, and the ensemble is
stochastically symmetric under such transformations. \linebreak
Equation (\ref{eq_prandom}) indicates that, for each of the
$N(N-1)/2$ pairs of vertices, the probability of an undirected link
being there is $p$. The probability of exactly $L(G)$ realised edges
is  $p^{L(G)}$ multiplied by the probability $(1-p)^{N(N-1)/2-L(G)}$
of the complementary number $N(N-1)/2-L(G)$ of missing edges. This
case is therefore equivalent the Erd\H{o}s-R\'enyi random graph
model that we already mentioned in Section \ref{sec_levels}, in
which each edge is drawn, independently of each other, with
probability $p$. The entropy of the ensemble now depends on $p$, and
one can easily see that if $p=1/2$, Equation (\ref{eq_maxent}) is
recovered. Indeed, this is the case where each edge is equally
likely to be present and absent, which is another way to say that no
constraint has been enforced and the entropy is maximum. By
contrast, in the two cases $p=0$ and $p=1$ the entropy is $S=0$ as
there is no uncertainty about the resulting structure of the
network. Indeed, in these cases the ensemble completely shrinks to
the only possible network, {\em i.e.}, the empty graph and the
complete graph respectively. If one wants to use the random graph
model as a null model of a real network $G^*$, the maximum
likelihood principle applied to Equation (\ref{eq_prandom})
indicates \cite{mylikelihood} that the parameter $p$ must be set to
the value
\begin{equation}
p^*=\frac{2L(G^*)}{N(N-1)}
\label{eq_likelihoodp}
\end{equation}
which ensures that the expected number of links $\langle L\rangle$,
as defined by Equation (\ref{eq_X}), reproduces the number of links
$L(G^*)$ of that particular network:
\begin{equation}
\langle L\rangle=p^*\frac{N(N-1)}{2}=L(G^*)
\label{eq_likelihoodL}
\end{equation}
In the random graph model, the expected degree distribution is
binomial (in the large network limit with fixed average degree,
Poissonian) with mean $p^*(N-1)=2L(G^*)/N$. The failure of the
random graph model in reproducing the properties of real networks,
according to our discussion in Section \ref{sec_levels}, can then be
restated as the inefficacy of specifying the number of links as the
only property of a network. This also means that real networks are
generally not stochastically symmetric under transformations
preserving the total number of links. A less trivial choice is the
so-called \emph{configuration model} \cite{maslov,configuration}.
Assuming we are still interested in undirected binary networks, the
configuration model is a maximally random graph ensemble where the
degrees of all vertices, {\em i.e.} the \emph{degree sequence}
$\{k_i\}$, are specified. Note that, in terms of the adjacency
matrix $A$ of the graph,  the degree of vertex $i$ is $k_i=\sum_j
a_{ij}$, and the total number of links is twice the sum of the
degrees of all vertices: $L=\sum_{i<j}a_{ij}=\sum_i k_i/2$.
Therefore specifying the degree sequence automatically fixes also
the total number of links, which confirms that this model is more
constraining than the random graph one. The configuration model
naturally comes into play in the problem we described in Section
\ref{sec_statistical}, when we stressed the importance of comparing
a real network to a null model in order to separate genuine
higher-order correlations from mere effects of low-level
constraints. The degree sequence is an important constraint to
consider, because the widespread occurrence of scale-free
architectures implies that major topological differences across real
networks must be looked for in other properties beyond the degree
distribution. Note that specifying the degree sequence $\{k_i\}$ is
different from specifying the degree distribution $P(k)$. A given
degree sequence generates a unique degree distribution, but there
are many degree sequences ($N!$ permutations) generating the same
degree distribution. Therefore fixing the degree distribution is
less informative than specifying the entire degree sequence, and we
do not consider it here. For directed graphs, the configuration
model is naturally extended by simultaneously considering as
constraints the number of incoming links (\emph{in-degree}) and the
number of outgoing links (\emph{out-degree}) of all vertices.
Similarly, for weighted networks the constraints become the
\emph{strength} (total edge weight) of all vertices (the
\emph{strength sequence}), or the corresponding directed quantities
when applicable.

In the binary undirected case, the Hamiltonian of the configuration
model contains the degrees of all vertices: \be H(G)=\sum_{i=1}^N
\theta_i k_i(G) \label{eq_Hconf} \ee and it can be shown
\cite{newman_origin} that the form of $P(G)$  determined by the
above choice is
\begin{equation}
P(G)=\prod_{i<j}p_{ij}^{a_{ij}(G)}(1-p_{ij})^{1-a_{ij}(G)}=\frac{\prod_i x_i^{k_i(G)}}{\prod_{i<j}(1+x_i x_j)}
\label{eq_Pconf}
\end{equation}
where
\begin{equation}
p_{ij}=\frac{x_i x_j}{1+x_i x_j}
\label{eq_pij}
\end{equation}
and $x_i\equiv e^{-\theta_i}$ is another way to write the Lagrange
multiplier associated to $k_i$. In this model, edges are still
independent, but have different probabilities.

\begin{figure}[h]
\begin{center}
\includegraphics[width=0.6\textwidth]{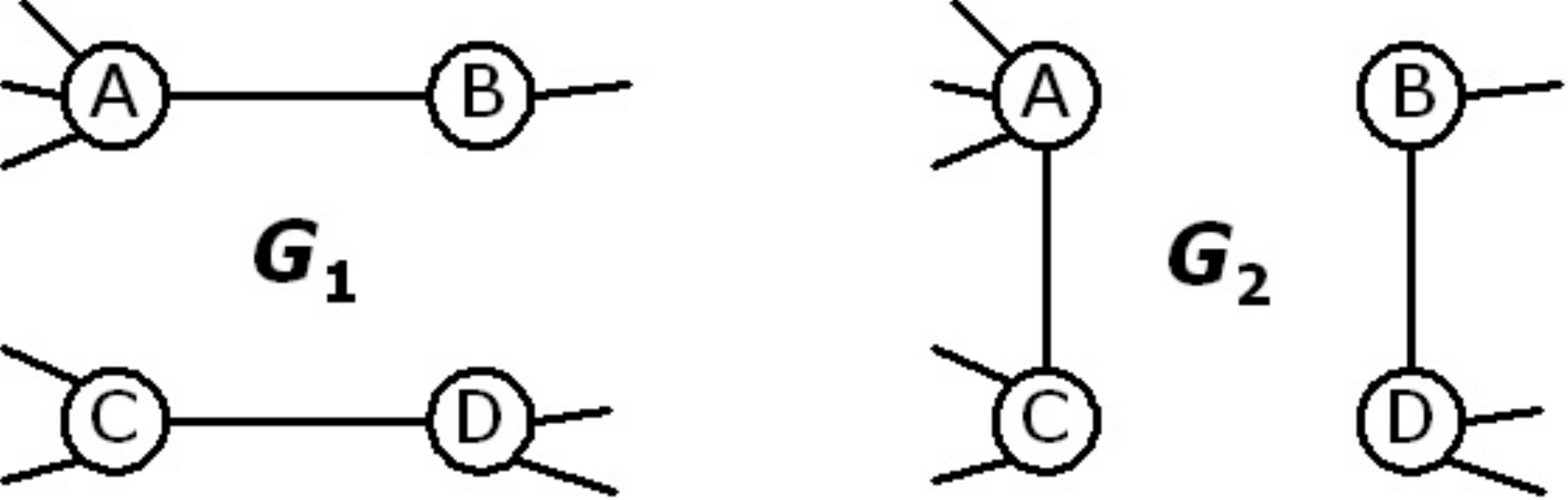}
\end{center}
\caption{The two undirected graphs $G_1$ and $G_2$ are identical,
except for the two pairs of edges shown. In the configuration model,
$G_1$ and $G_2$ occur with the same probability since their degree
sequences are the same. Reference \cite{maslov} exploits this
property as a recipe to iteratively randomize a real network while
preserving its degree sequence: in an elementary step, a graph like
$G_1$ is transformed into the graph $G_2$ (\emph{local rewiring
algorithm}). \label{fig_equiundgraphs}}
\end{figure}

The probability $P(G)$ of a graph $G$ only depends on its degree
sequence, as evident from \linebreak Equation (\ref{eq_Pconf}). Thus
any two graphs $G_1$ and $G_2$ with the same degree sequence
$\{k_i(G_1)\}=\{k_i(G_2)\}$ are equiprobable in the ensemble
specified by Equation (\ref{eq_Hconf}). A consequence of this
property is illustrated in Figure \ref{fig_equiundgraphs}, where we
show two  graphs $G_1$ and $G_2$ that have exactly the same
topology, except for the two edges shown. Graph $G_2$ can be
obtained from $G_1$ by replacing the two edges $(A-B)$ and $(C-D)$
with the two edges $(A-C)$ and $(B-D)$. Since this transformation
preserves the degree sequence, it is a symmetry of the Hamiltonian
defined in Equation (\ref{eq_Hconf}) and connects equiprobable
graphs. According to our definition in Section
\ref{sec_transformations}, the ensemble is stochastically symmetric
under such transformation. The equivalence classes of this symmetry
are sets of graphs with the same degree sequence.

This property has been used to constructively define an algorithm
that randomises a real network $G^*$ by iteratively selecting a pair
of edges and swapping the end-point vertices exactly as in
\linebreak Figure \ref{fig_equiundgraphs} \cite{maslov}. This
procedure, known as the \emph{local rewiring algorithm}, ergodically
explores the equivalence class where the real network $G^*$ belongs.
Any topological property of interest can be averaged across the set
of graphs produced by the algorithm and compared  with the value of
the same property in the original graph $G^*$. This allows to check
the effects of the degree sequence alone on the other topological
properties. As we mentioned, this null model is restricted to only
one equivalence class of the symmetry (it is a \emph{microcanonical
ensemble}), and requires that averages are numerically performed
over the graphs sampled by the local rewiring algorithm. By
contrast, the null model defined by Equation (\ref{eq_Hconf})
explores the entire set of $2^{N(N-1)/2}$ undirected graphs (it is a
\emph{(grand)canonical ensemble}), and allows to obtain the
expectation values analytically through Equation (\ref{eq_X}). This
requires that the parameters $\{x_1,\dots,x_N\}$ are set to the
values $\{x^*_1,\dots,x^*_N\}$ that maximise the likelihood to
obtain the real network $G^*$ \cite{mylikelihood,myrandomization}.
These values are found by solving the following $N$ coupled
equations
\begin{equation}
\langle k_i\rangle=\sum_{j\ne i}\frac{x^*_i x^*_j}{1+x^*_i x^*_j}=k_i(G^*)\qquad \forall i
\label{eq_likelihoodk}
\end{equation}
ensuring that the expected degree sequence coincides with the
observed one, and thus generalising Equation (\ref{eq_likelihoodL}).
As we already anticipated in Section \ref{sec_statistical}, an
important conclusion drawn from the analysis of the configuration
model is that, if real-world scale-free degree distributions are
specified, higher-order patterns are automatically generated. In
particular, the average nearest neighbour degree and the clustering
coefficient of a vertex with degree $k$ are both found to decrease
with $k$ \cite{maslov,newman_origin,myrandomization}. These patterns
should not be interpreted necessarily as the result of additional
mechanisms, beyond those required to explain the form of the degree
distribution. Note that if a real network is found to be well
reproduced by the configuration model, then it is stochastically
symmetric under transformations preserving the degree sequence. Also
note that any two vertices $i$ and $j$ with the same degree
$k_i(G^*)=k_j(G^*)$ in the real network are statistically equivalent
in the sense specified in Section \ref{sec_statistical}. This is
because Equation (\ref{eq_likelihoodk}) implies that those vertices
would be assigned the same parameter value $x_i^*=x^*_j$, and would
therefore have the same expected topological properties as discussed
for the fitness model in Section \ref{sec_properties}. Whereas
permutations of structurally equivalent vertices lead to exactly the
same topology and are therefore automorphisms (exact symmetries) of
the network, permutations of statistically equivalent vertices
(here, vertices with the same degree) are stochastic symmetries of
the network, if the latter is in accordance with the configuration
model. This is an interesting and important relation between
ensemble equiprobability,  symmetry under permutation of vertex
properties, and statistical equivalence. If the ensemble is not a
good model of the real network, which signals the presence of
mechanisms that break the postulated equiprobability symmetry, then
the real network is not stochastically symmetric under
transformations preserving the degree sequence, and vertices
displaying the same values of the enforced constraints are no longer
statistical equivalent.

Note that Equation (\ref{eq_Pconf}) generalises Equation
(\ref{eq_prandom}), and also that Equation (\ref{eq_pij}) can be
viewed as a particular case of the connection probability
$p(x_i,x_j)$ introduced in the fitness model we described in Section
\ref{sec_properties}. Indeed, the configuration model and the
fitness model both reduce to the random graph case if $x_i=x_0$
$\forall i$, {\em i.e.} if all vertices have the same properties. In
this case, the entropy associated with Equation (\ref{eq_pij})
coincides with the one associated with Equation (\ref{eq_prandom}).
By contrast, if the $x_i$'s are heterogeneously distributed, the
entropy is significantly decreased. In particular, the values of the
$x_i$'s required in order to enforce a scale-free degree
distribution as observed in real networks are approximately
power-law distributed, a result implying a strong reduction of the
entropy of the ensemble associated with the degree sequence of real
networks. In particular, it was shown that networks with degree
distribution $P(k)\propto k^{-2}$ have remarkably small entropy
\cite{ginestra_entropy} and can be generated deterministically
\cite{fitness} like regular graphs. We therefore see that network
complexity, as signalled in this example by a scale-free degree
distribution, can lead to a decrease in the stochastic symmetry
associated with ensemble equiprobability, and to a substantial
decrease in the corresponding entropy. From the perspective of the
amount of information required in order to reproduce them, real
networks (and possibly many real complex systems) turn out to
achieve an unsuspected degree of order by following a nontrivial
path, which is completely different from that taken by regular
structures.

\subsection{Symmetry under Network Partitioning: Modularity and Communities\label{sec_communities}}
\begin{figure}[h]
\begin{center}
\includegraphics[width=0.6\textwidth]{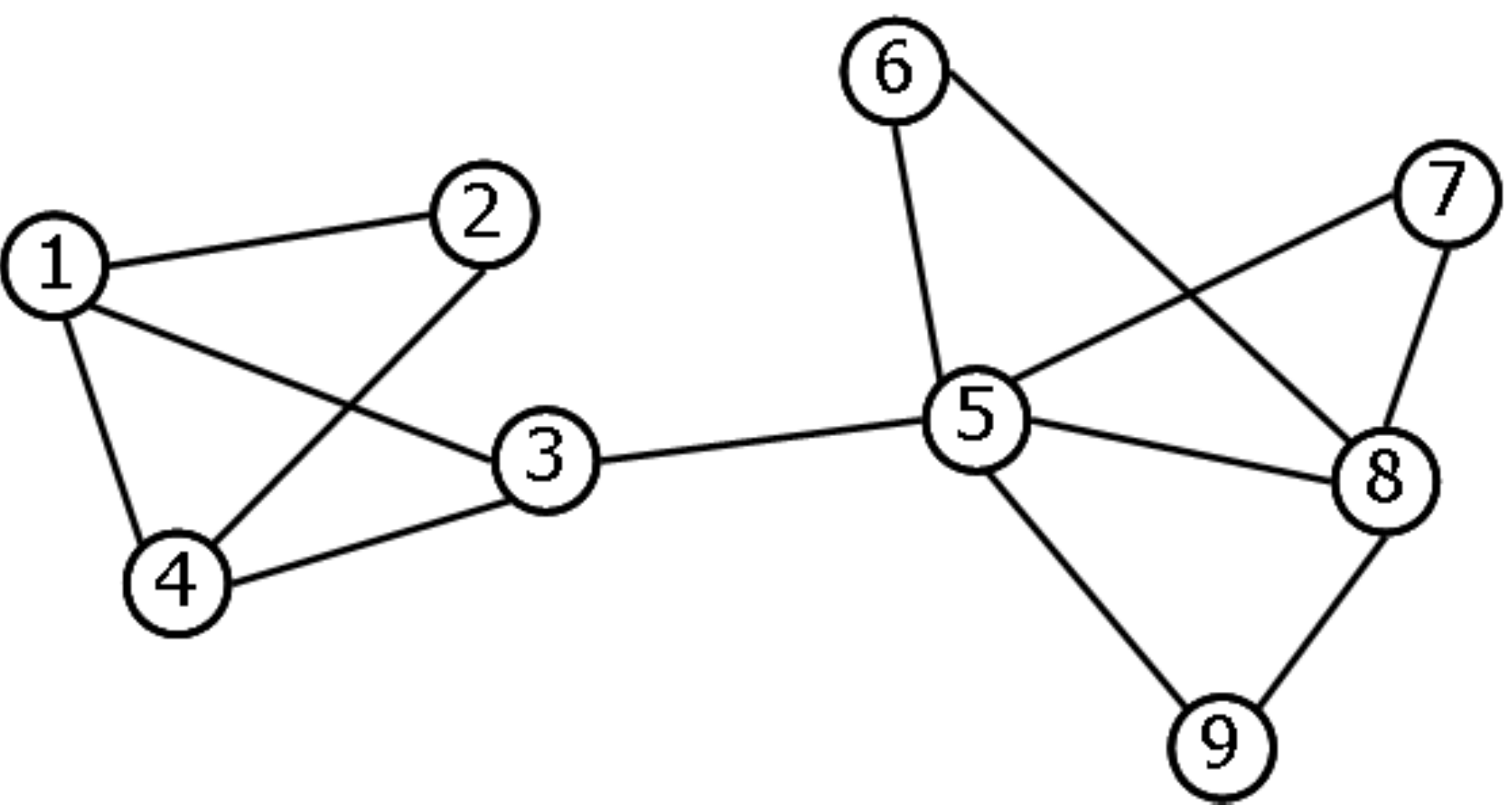}
\end{center}
\caption{Example of an undirected network with $N=9$ vertices, that
can be clearly grouped into $2$ non-overlapping communities:
vertices $1$ to $4$ form one community, and vertices $5$ to $9$ form
a second community. Intra-community links are denser than
inter-community ones. \label{fig_communities}} \vspace{-0.8cm}
\end{figure}

As we briefly mentioned in Section \ref{sec_levels}, real networks
are found to display inhomogeneous link density, and to be
partitioned into \emph{communities} of vertices
\cite{santo_communities}. Several different definitions of a
community have been introduced. Generally, these definitions try to
capture different aspects of the same simple idea: that communities
are more densely connected internally than with other communities,
so that intra-community links are typically denser than
inter-community ones. An example is shown in Figure
\ref{fig_communities} to illustrate this concept. This simple idea
can however give rise to technical difficulties when  implemented
into community detection algorithms and applied to large networks,
and as a result different methods have been developed, each dealing
with a different aspect of the problem. For instance, some methods
try to identify the \emph{optimal partition} of vertices into
non-overlapping subsets representing communities; others recognise
that the optimality of a partition depends on the resolution
adopted, and give a \emph{multi-resolution} output where communities
are hierarchically nested into each other; others are devised to
identify \emph{overlapping} communities, {\em etc}. Presenting the
subtleties and diversity of the community detection problem is
beyond the scope of the present review, and the interested reader is
referred to the relevant literature \cite{santo_communities}. We
simply note here that the community structure of a network is
connected to a particular type of symmetry: the invariance under
network partitioning. To illustrate this idea, we consider as an
example a widely used quantity that measures the goodness of a
partition of a real undirected network into non-overlapping
communities, {\em i.e.} the \emph{modularity}
\begin{equation}
Q\equiv\frac{1}{L}\sum_{i<j} (a_{ij}-p_{ij})c_{ij}
\end{equation}
In the above definition, $a_{ij}$ is the entry of the adjacency
matrix $A$ of the real network, $L=\sum_{i<j}a_{ij}$ is the observed
number of links, $p_{ij}$ is the probability that vertices $i$ and
$j$ are connected under a null model chosen as a reference, and
$c_{ij}$ indicates if in the partition under consideration vertices
$i$ and $j$ are placed in the same community ($c_{ij}=1$) or in
different communities ($c_{ij}=0$). Typically, the null model
considered is the configuration model (see Section
\ref{sec_equiprobability}). Since different partitions of the same
network correspond to different sets of values $\{c_{ij}\}$, $Q$ can
be used to assess the performance of a partition in correctly
placing in the same community ($c_{ij}=1$) pairs of vertices that
are connected ($a_{ij}=1$) despite the null model predicts a low
connection probability ($p_{ij}\approx 0$), and in placing in
different communities ($c_{ij}=0$) pairs of vertices that are not
connected ($a_{ij}=0$) despite the null model predicts a high
connection probability ($p_{ij}\approx 1$). Larger values of $Q$
represent better partitions. If the network is well reproduced by
the null model, then one expects a value of $Q$ close to zero,
independently of the partition. To see this, imagine that the
network has indeed been generated by the null model. If several
realisations of the network are generated, then the expected value
pf $a_{ij}$ is $p_{ij}$ and the expected modularity is \be \langle
Q\rangle=0 \ee independently of $c_{ij}$. This means that a network
with no community structure is stochastically invariant (in the
sense specified in Section \ref{sec_transformations}) under vertex
partitioning, as all reassignments of vertices to different
communities preserve on average the modularity. The modular
structure of real networks can be therefore seen as a
symmetry-breaking property. In some networks, the maximisation of
the modularity can be very complicated numerically, as there are
many competing partitions with similar values of $Q$
(computationally, finding the partition corresponding to the global
maximum of $Q$ is a  NP-hard problem). This indicates that in real
networks the overall invariance under partitioning is often broken
down to equivalence classes containing partitions with approximately
equal modularity.

\subsection{Edge Weight Permutation Invariance\label{sec_weights}}
As the last example of symmetries in networks, we consider an
invariance that naturally comes into play in the analysis of
weighted networks. Weighted networks are described by a non-negative
matrix $W$ rather than by a binary adjacency matrix $A$. The entry
$w_{ij}$ of the matrix $W$ represents the weight of the edge from
vertex $i$ to vertex $j$ (if $w_{ij}=0$ no edge is there). In the
analysis of weighted networks, a crucial point is assessing whether
the knowledge of edge weights indeed conveys additional information
with respect to the knowledge of the binary topology. This problem
has been tackled by introducing suitable definitions of structural
properties that make explicit use of the empirical edge weights and
that distinguish between the real network and suitably randomised
counterparts \cite{vespy_weighted,myensemble,kertesz_clustering}.

\begin{figure}[h]
\begin{center}
\includegraphics[width=0.8\textwidth]{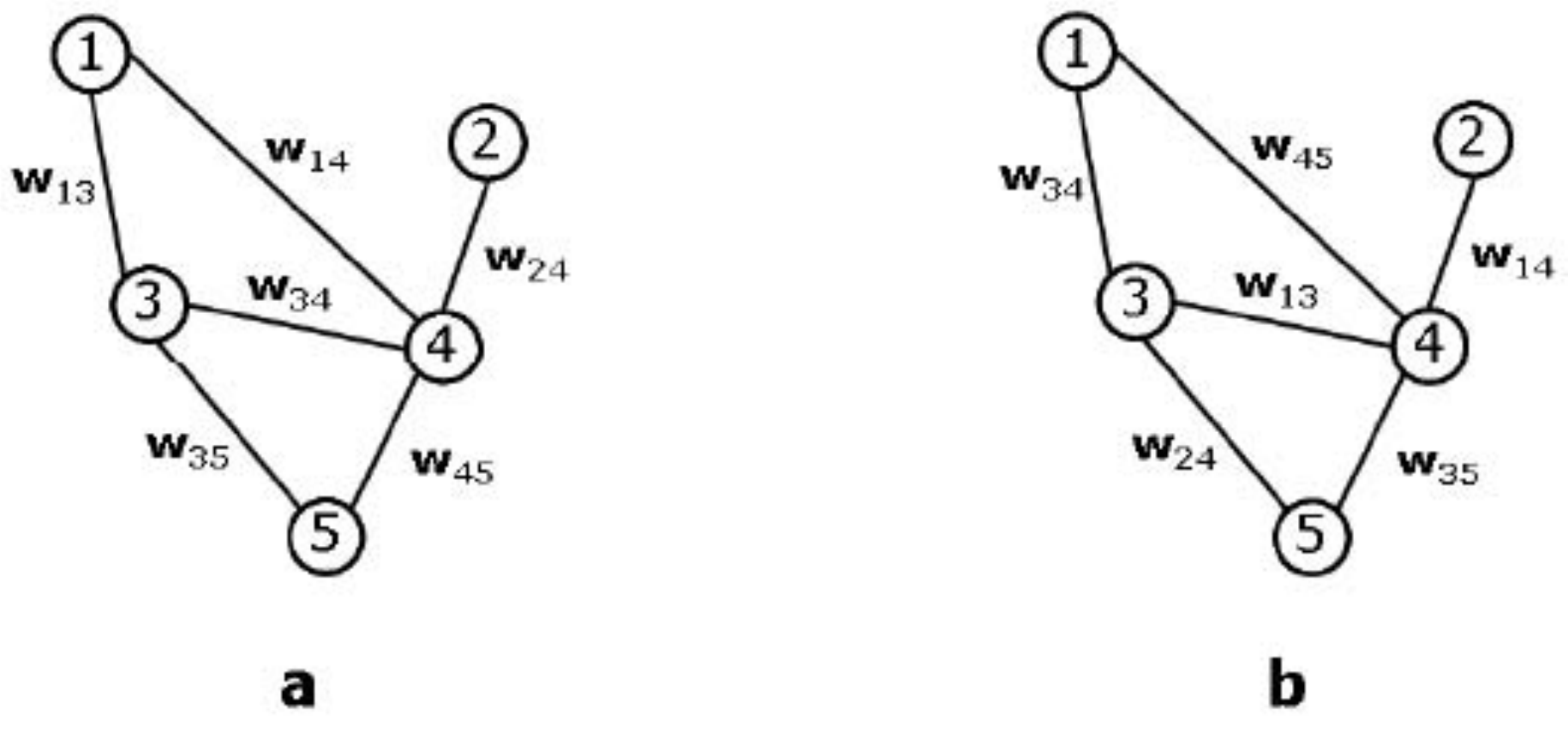}
\end{center}
\caption{Construction of a null model, alternative to the weighted
generalization of the local rewiring algorithm defined in Figure
\ref{fig_equiundgraphs}, against which the properties of a real
weighted network can be compared. (a) A real network is considered,
where each link $(i-j)$ has an observed weight $w_{ij}$. (b) The
empirical weights $\{w_{ij}\}$ are randomly shuffled across the
links of the network, which are kept in the original positions (the
topology is unchanged). Iterating this procedure generates an
ensemble of randomized weighted networks. In such a way, the
correlations between weights and topology are removed, and one has a
family of uncorrelated benchmarks for the empirical network.
\label{fig_weights}}
\end{figure}

The randomised case can be either a weighted generalisation of the
maximally random networks described in Section
\ref{sec_equiprobability} \cite{mybosefermi}, or a different null
model providing a reference where weights and topology are
uncorrelated, so that weighted properties reduce to simpler binary
properties \cite{vespy_weighted}. The latter null model consists in
taking the real network, keeping its topology fixed, and randomly
reshuffling the values of the weights across the edges (see Figure
\ref{fig_weights}).

Iterating this procedure generates an ensemble of randomised
networks where any correlation existing between weights and topology
is destroyed.  This provides a reference for the analysis of the
original real network. A prototypical example of the deviation of
real networks from the uncorrelated case is the generally observed
power-law relation between the degree $k_i=\sum_{j\ne i} a_{ij}$ and
the strength $s_i=\sum_{j\ne i} w_{ij}$ of vertices: \be s_i\propto
k_i^\beta \ee where usually $\beta>1$. By contrast, in the
uncorrelated case provided by the null model, the strength is simply
proportional to the degree, which is its unweighted counterpart.
This yields $\beta=1$. Similar results are found for other
quantities. In general, if suitable weighted structural properties
are defined and averaged across the uncorrelated ensemble, the
output is in a trivial relation with the purely binary counterparts
of these properties \cite{vespy_weighted}.

We note that the above problem can be rephrased as a generalisation
of the symmetry we introduced in Section \ref{sec_properties}.
Indeed, weights can be considered as non-topological properties
attached to edges (rather than to vertices). Nontrivial correlations
between weights and topology correspond to a lack of invariance of
the real network under permutations of weights across the edges.
Whereas uncorrelated weighted networks are stochastically symmetric
under such permutations, real networks are found to display strong
correlations. Therefore, we find again that network complexity, now
at the level of weights, can manifest itself in terms of
symmetry-breaking correlations restricting possible network
invariances to smaller equivalence classes.

\section{Conclusions\label{sec_conclusions}}
In this review we have discussed various types of symmetries
encountered in the analysis of real networks. Symmetry concepts turn
out to offer an insightful review of network theory from an unusual
perspective. In particular, we have shown that many empirical
properties of complex networks can be rephrased in terms of (the
lack of) exact or stochastic symmetries. Exact symmetries of a
network are transformations that map the network onto itself. If
such transformations are permutations of vertices, they are the
automorphisms of the graph. Special cases include symmetries induced
by structural equivalence (Section \ref{sec_structural}) or by an
embedding of vertices in some space, such as translational symmetry
(Section \ref {sec_translational}). Stochastic symmetries of a
network are transformations that map the network onto a different
one in the same statistical ensemble, and are therefore associated
with a family of graphs with similar properties, rather than with a
single graph. We have discussed stochastic vertex permutation
symmetries in the context of statistical equivalence (Section
\ref{sec_statistical}) and invariance under permutation of vertex
properties (Section \ref{sec_properties}). We have also discussed
transformations not associated with permutations of vertices, such
as scale invariance (Section \ref{sec_scaleinvariance}), ensemble
equiprobability (Section \ref {sec_equiprobability}), invariance
under vertex partitions (Section \ref{sec_communities}), and edge
weight permutations (Section \ref{sec_weights}). We have shown that
various correlation patterns observed in real networks imply that
the above symmetries only hold within disjoint equivalence classes,
specified for instance by some property of vertices. This often
indicates which are the most informative topological properties of
real networks: those that partition vertices (or other parts of the
graph) into the equivalence classes of some (stochastic) symmetry.
Therefore we believe that the study of symmetry in networks is a
promising field of research, which deserves more attention in future
investigations. While automorphism groups are well studied within
discrete mathematics for particular classes of graphs generated
according to deterministic rules, the analysis of symmetry in real
heterogeneous networks is far less developed. We suggested that real
networks---as any real entity characterized by imperfections or
errors---necessarily require a stochastic notion of symmetry. Our
preliminary investigation shows that such an expanded scenario may
lead to very informative results, as it can detect ordered patterns
in intrinsically noisy contexts, where exact techniques fail. In the
companion paper \cite{symmetry2}, we apply our ideas in more detail
and show the full power of stochastic symmetry in a particular case.

\section*{Acknowledgements}
D.G. acknowledges financial support from the European Commission 6th
FP (Contract CIT3-CT-2005-513396), Project: DIME - Dynamics of
Institutions and Markets in Europe.

\bibliographystyle{mdpi}
\makeatletter
\renewcommand\@biblabel[1]{#1. }
\makeatother

\begin{thebibliography}{100}
\bibitem{symmetry2}
Ruzzenenti, F.; Garlaschelli, D.; Basosi, R.
Complex networks and symmetry II: Reciprocity and evolution of world trade.
{\em Symmetry} {\bf 2010}, {\em 2}, X--X.

\bibitem{guidosbook}
Caldarelli, G. {\em Scale-Free Networks: Complex Webs in Nature and Technology}; Oxford University Press: Oxford, UK, 2007.

\bibitem{largescalestructure}
Caldarelli, G.; Vespignani, A. {\em Large Scale Structure and Dynamics of Complex Networks}; World Scientific Press: Singapore, 2007.

\bibitem{dynamicalprocessesoncomplexnetworks}
Barrat, A.; Barthelemy, M.; Vespignani, A. {\em Dynamical Processes on Complex Networks}; Cambridge University Press: New York, NY, USA, 2008.

\bibitem{ecologicalnetworks}
Pascual, M.; Dunne, J.A. {\em Ecological Networks: Linking Structure to Dynamics in Food Webs}; Oxford University Press: New York, NY, USA, 2006.

\bibitem{internet}
Pastor-Satorras, R.; Vespignani, A. {\em Evolution and Structure of the Internet: A Statistical Physics Approach}; Cambridge University Press: Cambridge, UK, 2004.

\bibitem{networksincellbiology}
Buchanan, M.; Caldarelli, G.; De Los Rios, P.; Rao, F.; Vendruscolo, M. {\em Networks in Cell Biology}; Cambridge University Press: Cambridge, UK, 2010.

\bibitem{adaptivenetworks}
Gross, T.; Sayama, H. {\em Adaptive Networks}; Springer/NECSI: Cambridge, Massachusetts (USA), 2009.

\bibitem{harary}
Harary, F. {\em Graph Theory}; Addison-Wesley: Reading, MA, USA, 1994.

\bibitem{vespy_weighted}
Barrat, A.; Barthelemy, M.; Pastor-Satorras, R.; Vespignani, A. The architecture of complex weighted networks. {\em PNAS} {\bf 2004}, {\em 101}, 3747--3752.

\bibitem{myensemble}
Ahnert, S.E.; Garlaschelli, D.; Fink, T.M.A.; Caldarelli, G. Ensemble approach to the analysis of weighted networks. {\em Phys. Rev. E} {\bf 2007}, {\em 76}, 016101.

\bibitem{kertesz_clustering}
Saramaki, J.; Kivela, M.; Onnela, J.-P.; Kaski, K.; Kertesz, J. Generalizations of the clustering coefficient to weighted complex networks. {\em Phys. Rev. E} {\bf 2007}, {\em 75}, 027105.

\bibitem{myreciprocity}
Garlaschelli, D.; Loffredo, M.I. Patterns of link reciprocity in directed networks. {\em Phys. Rev. Lett.} {\bf 2004}, {\em 93}, 268701.

\bibitem{mymultispecies}
Garlaschelli, D.; Loffredo, M.I. Multispecies grand-canonical models for networks with reciprocity. {\em Phys. Rev. E} {\bf 2006}, {\em 73}, 015101.

\bibitem{giorgioclustering}
Fagiolo, G. Clustering in complex directed networks. {\em Phys. Rev. E} {\bf 2007}, {\em 76}, 026107.

\bibitem{myselforganized}
Garlaschelli, D.; Capocci, A.; Caldarelli, G. Self-organized network evolution coupled to extremal dynamics. {\em Nat. Physics} {\bf 2007}, {\em 3}, 813--817.

\bibitem{symmetry}
MacArthur, B.D.; S\'anchez-Garc\'ia, R.J.; Anderson, J.W.
Symmetry in complex networks. {\em Discrete Appl. Math.} {\bf 2008}, {\em 156}, 3525--3531.

\bibitem{quotient}
Xiao, Y.; MacArthur, B.D.; Wang, H.; Xiong, M.; Wang, W.
Network quotients: Structural skeletons of complex systems.
{\em Phys. Rev. E} {\bf 2008}, {\em 78}, 046102.

\bibitem{redundancy}
MacArthur, B.D.; S\'anchez-Garc\'ia, R.J.
Spectral characteristics of network redundancy.
{\em Phys. Rev. E} {\textbf 2009}, {\em 80}, 026117.

\bibitem{symmetry_wtw}
Wang, H.; Yan, G.; Xiao, Y.
Symmetry in world trade network.
{\em J. Syst. Sci. Complex.} {\bf 2009}, {\em 22}, 280--290.

\bibitem{smallworld}
Watts, D.J.; Strogatz, S.H. Collective dynamics of `small-world' networks. {\em Nature} {\bf 1998}, {\em 393}, 440--442.

\bibitem{powerlaws}
Newman, M.E.J. Power laws, Pareto distributions and Zipf's law. {\em Contemp. Phys.} {\bf 2005}, {\em 46}, 323--351.

\bibitem{shlomo}
Song, C.; Havlin, S.; Makse, H.A. Self-similarity of complex networks. {\em Nature} {\bf 2005}, {\em 433}, 392--395.

\bibitem{renormalization}
Stanley, H.E. Scaling, universality, and renormalization: Three pillars of modern critical phenomena. {\em Rev. Mod. Phys.} {\bf 1999}, {\em 71}, S358--S366.

\bibitem{BA}
Barabasi, A.-L.; Albert, R. Emergence of Scaling in Random Networks. {\em Science} {\bf 1999}, {\em 286}, 509--512.

\bibitem{fitness}
Caldarelli, G.; Capocci, A.; De Los Rios, P.; Munoz M.A. Scale-free networks from varying vertex intrinsic fitness. {\em Phys. Rev. Lett.} {\bf 2002}, {\em 89}, 258702.

\bibitem{wasserman}
Wasserman, S.; Faust, K. {\em Social Network Analysis: Methods and Applications}; Cambridge University Press: New York, NY, USA, 1994.

\bibitem{myfoodwebs}
Garlaschelli, D.; Caldarelli, G.; Pietronero, L. Universal scaling relations in food webs. {\em Nature} {\bf 2003}, {\em 423}, 165--168.

\bibitem{motifs}
Alon, U.
Network motifs: Theory and experimental approaches.
{\em Nat. Rev. Genet.} {\bf 2007}, {\em 8}, 450--461.

\bibitem{newman_assortative}
Newman, M.E.J. Mixing patterns in networks. {\em Phys. Rev. E} {\bf 2003}, {\em 67}, 026126.

\bibitem{hierarchy}
Ravasz, E.; Barabasi, A.-L. Hierarchical organization in complex networks. {\em Phys. Rev. E} {\bf 2003}, {\em 67}, 026112.

\bibitem{maslov}
Maslov, S.; Sneppen, K.; Zaliznyak, A. Detection of topological patterns in complex networks: correlation profile of the Internet. {\em Physica A} {\bf 2004}, {\em 333}, 529--540.

\bibitem{newman_origin}
Park, J.; Newman, M.E.J. Origin of degree correlations in the Internet and other networks. {\em Phys. Rev. E} {\bf 2003}, {\em 68}, 026112.

\bibitem{pin_jackson}
Currarini, S.; Jackson, M.O.; Pin, P. Identifying the roles of race-based choice and chance in high school friendship network formation. {\em PNAS} {\bf 2010}, doi: 10.1073/pnas.0911793107.

\bibitem{pin_ginestra}
Bianconi, G.; Pin, P.; Marsili, M. Assessing the relevance of node features for network structure. {\em PNAS} {\bf 2009}, {\em 106}, 11433--11438.

\bibitem{mylikelihood}
Garlaschelli, D.; Loffredo, M.I. Maximum likelihood: Extracting unbiased information from complex networks. {\em Phys. Rev. E} {\bf 2008}, {\em 78}, 015101.

\bibitem{ramasco}
Ramasco, J.J.; Mungan, M. Inversion method for content-based networks. {\em Phys. Rev. E} {\bf 2008}, {\em 77}, 036122.

\bibitem{newman_statistical}
Park, J.; Newman, M.E.J. Statistical mechanics of networks. {\em Phys. Rev. E} {\bf 2004}, {\em 70}, 066117.

\bibitem{ginestra_entropy}
Bianconi, G. Entropy of network ensembles. {\em Phys. Rev. E} {\bf 2009}, {\em 79}, 036114.

\bibitem{mybosefermi}
Garlaschelli, D.; Loffredo, M.I. Generalized Bose-Fermi Statistics and Structural Correlations in Weighted Networks. {\em Phys. Rev. Lett.} {\bf 2009}, {\em 102}, 038701.

\bibitem{configuration}
Newman, M.E.J.; Strogatz, S.H.; Watts, D.J. Random graphs with arbitrary degree distributions and their applications. {\em Phys. Rev. E} {\bf 2001}, {\em 64}, 026118.

\bibitem{myrandomization}
Squartini, T.; Garlaschelli, D. Exact maximum-likelihood method to
detect patterns in real networks. Preprint available at
http://arxiv.org/

\bibitem{santo_communities}
Fortunato, S. Community detection in graphs. {\em Physics Reports} {\bf 2010}, {\em 486}, 75--174.


\end{thebibliography}

\end{document}